\documentclass[a4paper,notitlepage,11pt]{article}
\pdfoutput=1
\usepackage[utf8]{inputenc}
\usepackage{geometry}
\usepackage[english]{babel}
\usepackage{amsmath, amsfonts, graphicx}
\usepackage{mathtools}
\usepackage{graphicx}
\usepackage{float}
\usepackage{hyperref}
\usepackage[labelfont=bf,font={small}]{caption}
\usepackage{subcaption}
\usepackage{tikz}
\usepackage{footmisc}
\usepackage{multirow}
\usepackage{url}
\usepackage{cleveref}
\usepackage{enumerate}
\usepackage{physics}
\usepackage{bm}
\usepackage{epstopdf}
\numberwithin{equation}{section}
\allowdisplaybreaks

\usetikzlibrary{calc}

\newcounter{markeq}
\title{\textbf{An Investigation of AdS$_{2}$ Backreaction and Holography}}
\author{\small \textbf{Julius Engelsöy}\thanks{\href{mailto:engelsoy@kth.se}{\protect\path{engelsoy@kth.se}}, \href{mailto:engelsoy@princeton.edu}{\protect\path{engelsoy@princeton.edu}}}\vspace{0.2cm}\\
\begin{tabular}[H]{c c}
\small \textit{Physics Department,} & \small \textit{Physics Department,}\\
\small \textit{Royal Institute of Technology,} & \small \textit{Princeton University,}\\
\small\textit{114 28 Stockholm, Sweden} & \small\textit{NJ 08544, USA}
\end{tabular}
\vspace{0.5cm}
\\
\small \textbf{Thomas G. Mertens}\thanks{\href{mailto:tmertens@princeton.edu}{\protect\path{tmertens@princeton.edu}}}\vspace{0.2cm}\\
\small \textit{Physics Department,}\\
\small \textit{Princeton University,}\\ \small\textit{NJ 08544, USA}
\vspace{0.5cm}
\\
\small \textbf{Herman Verlinde}\thanks{\href{mailto:verlinde@princeton.edu}{\protect\path{verlinde@princeton.edu}}}\vspace{0.2cm}\\
\small \textit{Physics Department and Princeton Center for Theoretical Science,}\\
\small \textit{Princeton University,}\\ \small\textit{NJ 08544, USA}}
\date{}
\begin{document}
\maketitle
\thispagestyle{empty}
\begin{abstract}
We investigate a dilaton gravity model in AdS$_{2}$ proposed by Almheiri and Polchinski \cite{Almheiri:2014cka} and develop a 1d effective description in terms of a dynamical boundary time with a Schwarzian derivative action. We show that the effective model is equivalent to a 1d version of Liouville theory, and investigate its dynamics and symmetries via a standard canonical framework. We include the coupling to arbitrary conformal matter and analyze the effective action in the presence of possible sources. We compute commutators of local operators at large time separation, and match the result with the time shift due to a gravitational shockwave interaction. We study a black hole evaporation process and comment on the role of entropy in this model.
\end{abstract}
\newpage
\tableofcontents
\thispagestyle{empty}
\newpage
\setcounter{page}{1}
\section{Introduction}
\def\sigmabp{}
\def\sigmabm{}
\def\is{\! & \! = \! & \!}

The microscopic dynamics that underlies the holographic AdS/CFT dictionary is still incompletely understood. Most of these questions persist in low dimensional examples. Two-dimensional anti-de Sitter space-times (AdS${}_{2}$) have proven to be remarkably delicate when it comes to understanding holography. One elementary reason is that any finite energy perturbation causes a significant backreaction on the AdS space-time \cite{Maldacena:1998uz}\cite{Strominger:1998yg}. It has also proven to be a hard problem to identify the dual quantum many body system, and to decide whether it should take the form of conformal quantum mechanics or a boundary 2d CFT \cite{Balasubramanian:2009bg}.

Recently, Almheiri and Polchinski have considered a model that in the IR regime reduces to pure AdS${}_2$, but in the UV gets adjusted by a non-trivial dilaton profile. This regulates the gravitational back-reaction, allowing one to set up a more meaningful holographic dictionary. The model was first proposed some time ago by Jackiw \cite{Jackiw:1984je}
 and Teitelboim	\cite{Teitelboim:1983ux}.
One of the interesting conclusions is that the time coordinate on the boundary becomes a dynamical variable. 

Quantum field theory interacting with 2d dilaton gravity has been studied extensively in earlier work as toy models that exhibit black hole formation and evaporation. For massless matter fields, conformal invariance provides powerful techniques for getting concise quantitative results \cite{Callan:1992rs}\cite{Giddings:1992ff}\cite{Spradlin:1999bn}\cite{Fabbri:2005mw}. 
Given that pure 2d gravity does not possess any local degrees of freedom, 
it is not surprising that all the gravitational dynamics can be encoded in the form of an effective  dynamical moving mirror model \cite{Chung:1993rf}\cite{Schoutens:1994st}. 
In this paper, we analyze whether the boundary dynamics of the Jackiw-Teitelboim or Almheiri-Polchinski (AP) model, studied in depth in \cite{Almheiri:2014cka}, can be described in a similar way.
We find that the 1d action that described the dynamics of the 1d boundary is given by the Schwarzian derivative. The same type of action was found to describe the dynamics of 
the dynamical time variable in the Sachdev-Ye-Kitaev (SYK) model \cite{Sachdev:1992fk}\cite{Kitaev}\cite{Sachdev:2015efa}\cite{Polchinski:2016xgd}\cite{Jevicki:2016bwu}\cite{Maldacena:2016hyu}.
The connection between the SYK models and AdS${}_2$ holography was first proposed in \cite{Sachdev:2010um}.

A hallmark feature shared by the SYK model and AP model is that early perturbations have an exponentially growing effect on late time observables, with a universal exponent proportional to temperature. In the dilaton gravity context,  this Lyapunov-type behavior arises due to shockwave interactions near the black hole horizon \cite{Dray:1984ha}. Their quantum effect has been extensively studied in earlier work \cite{Chung:1993rf}\cite{Schoutens:1994st}, where it was shown that it gives rise to exponentially growing commutators between early and late time observables.  Within a holographic context, the link between shockwave interactions in the bulk, and chaotic behavior of the boundary theory has been substantially studied in the recent years \cite{Shenker:2013pqa}\cite{Shenker:2013yza}\cite{Jackson:2014nla}\cite{Turiaci}\cite{Polchinski:2015cea}\cite{Maldacena:2015waa}. 

The AP model, or more generic systems dual to AdS$_2$ dilaton gravity, recently received quite some attention \cite{Jensen:2016pah}\cite{Maldacena:2016upp}. The results we report here provide a complimentary perspective.

This paper is structured as follows. Section \ref{sect2} contains a short review on the AP model itself. Section \ref{sect4} discusses the crucial features of this dilaton-gravity-matter model that we will need further on: the reparametrization equation of the boundary time coordinate in this model, and the computation of the holographic stress tensor expectation value. 
Section \ref{sect6} introduces our 1d effective model that captures the previous bulk features. We propose a model for a fluctuating boundary and then analyze how to couple this to a 2d CFT. We analyze the structure and symmetries present in the theory. Next, in Section \ref{commutatorsection}, we compute the commutator of local operators, relevant for OTO correlators, and find maximal chaotic behavior. 
Finally, Section \ref{sect3} introduces quantum effects in this model. We first develop effective non-local actions, relevant for computing correlators. Secondly, the expressions obtained in this work are applied to an interesting concrete example of relevance in the black hole information paradox context, where we model a 2d evaporating black hole, and describe the decrease of energy in the system as it evaporates through Hawking emission. 
We end with a brief conclusion in Section \ref{concl}. 
Some more detailed computations are contained in the appendices.

\def\beq{\begin{equation}}
\def\eeq{\end{equation}}
\def\bea{\begin{eqnarray}}
\def\eea{\end{eqnarray}}
\def\spc{\hspace{1pt}}
\def\smpc{\hspace{.5pt}}
\def\nspc{\hspace{-1pt}}
\def\nsmpc{\hspace{-.5pt}}

\def\ketw{
{{\kappa E} }}

\def\twek{
{\frac{E} \kappa }}

\def\pze{
{\frac{ {\pi_\tau + P_0}_{}} E^{}}}

\def\oops{
{\frac{1_{}} { \pi_\tau + P_0^{}} }}

\section{The Almheiri--Polchinski model}
\label{sect2}
In this section we give a brief review of the 2d dilaton gravity model introduced first by Jackiw \cite{Jackiw:1984je}
 and Teitelboim	\cite{Teitelboim:1983ux} and examined recently by Almheiri and Polchinski in \cite{Almheiri:2014cka}. Pure gravity in two dimensions has no dynamics. By adding a dilaton scalar field $\Phi$, one can write a dynamical action of the general form
\bea
S = \frac{1}{16\pi G}\int \dd^{2}x\sqrt{-g}\Bigl( \Phi^{2}R - V(\Phi) \Bigr) + S_{\rm matter} 
\,,\label{action}
\eea
where $S_{\rm matter}$ is some arbitrary matter system, coupled to the 2d metric  $\dd s^2 = g_{ab} \dd x^a \dd x^b$ and dilaton $\Phi$. The dilaton and metric are now dynamical, but have no local excitations: their value is uniquely fixed via their equation of motion, in terms of the energy momentum flow of the matter. Following \cite{Almheiri:2014cka}, we require that the classical background metric is given by AdS${}_2$ geometry, regardless of the presence of the matter. This requirement prescribes that the matter action does not depend on the dilaton, and fixes the dilaton potential term to be of the form
\bea
V(\Phi) \is 2 - 2\Phi^2 \, .
\eea
In conformal gauge $
\dd s^2 = 
-e^{2\omega(u,v)} \dd u \dd v,$
the equations of motion then take the form
\bea
4\partial_{u}\partial_{v}\omega + e^{2\omega} \is 0  \,,\label{metric}\\[1mm]
- e^{2\omega}\partial_{u}\left(e^{-2\omega}\partial_{u}\Phi^{2}\right)\is 8\pi G \, T_{uu} \,,\label{Tpp}\\[1mm]
- e^{2\omega}\partial_{v}\left(e^{-2\omega}\partial_{v}\Phi^{2}\right) \is 8\pi G \, T_{vv} \,,\label{Tmm}\\[1mm]
2\partial_{u}\partial_{v}\Phi^{2} + e^{2\omega}\left(\Phi^{2} - 1\right) \is 16 \pi G \, T_{uv} \,,\label{dilaton}
\eea
where $T_{ab}=   -\frac{2}{\sqrt{-g}}\frac{\delta {{S_{\rm matter}}_{\!}}_{}}{\delta g^{ab}}$  
denotes the matter energy-momentum tensor. 
The general solution to the metric equation of motion (\ref{metric}) is given by the AdS${}_2$ geometry
\bea
\dd s^2 \is -\frac{4\dd X^{+}\sigmabp \dd X^{-} \sigmabm }{(X^{+}\sigmabp  - X^{-}\sigmabm )^{2}}\,, 
\label{metricX}
\eea
where $X^\pm$ are general monotonic functions of the respective light-cone coordinates
\bea
\label{Xpar}
X^+  \equiv  X^{+}(u), \qquad \qquad X^- \equiv X^-(v)\, .
\eea
The AdS${}_2$ boundary is located at $X^+(u) = X^-(v)$.

In the following, we will assume that the matter sector is described by a conformal field theory. This means that, classically,
we can set $T_{uv}=0$. (We will discuss the modifications due to the trace anomaly in a later section.)
The equation of motion for the dilaton field $\Phi$ can then be explicitly integrated to
\begin{eqnarray}
\label{dilat}
\Phi^{2}(u,v) & = & 1 \spc +\spc  \frac{a }{X^{+}\nspc - X^{-}\sigmabm }\Bigl( 1 - \kappa \,( I_+ \spc +\spc I_- )\Bigr) \, , \\[2mm]
I_+(u,v) & = &   \int_{X^{+}\sigmabp}^{+\infty }\!\!\!\!\dd s\, {(s - X^{+})(s - X^{-}\sigmabp )}\, T_{++}(s)  \, ,\\[1mm]
I_ -(u,v) & = &  \int_{-\infty}^{X^{-}\sigmabm }\!\!\!\! \dd s\, (s - X^{+}\sigmabp )(s - X^{-}\sigmabm ) \, T_{--}(s) \,.
\end{eqnarray}
Here we introduced the integration constant $a$, and the short-hand notation 
 \bea
 \kappa \! \is \!  \frac{8\pi G}{a} \, .
 \eea
For physical reasons that will become apparent later, the integration constant $a$ needs to be nonzero and positive. 
We will treat it as a fixed parameter that specifies the asymptotic boundary condition of the dilaton field. As explained in \cite{Almheiri:2014cka}, 
it stabilizes the backreaction of the model under finite energy excitations. 
From the action \eqref{action} it is clear that the location at which $\Phi^{2} = 0$ is a strong coupling singularity. The choice $a>0$ prevents this singularity from reaching the boundary within a finite amount of time.

 \subsection{Black hole solution}
 
\begin{figure}[t]
\begin{center}
\medskip
\includegraphics[width=0.26\textwidth]{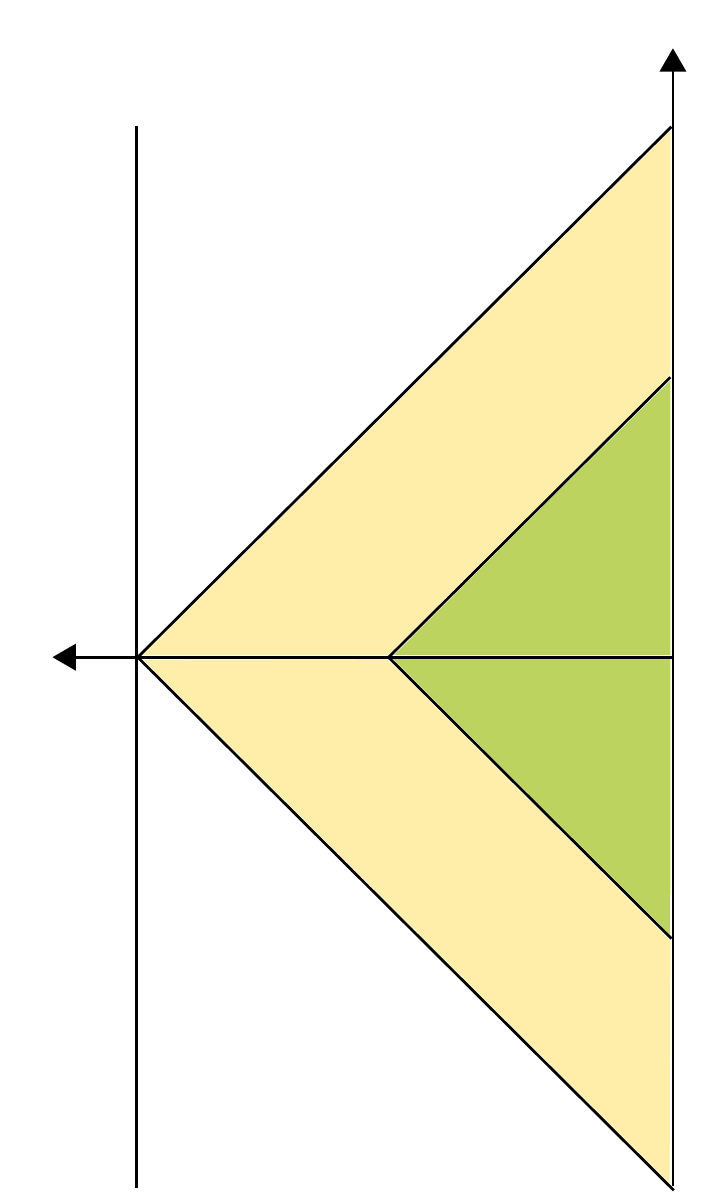}
\end{center}
\vspace{-2mm}
\caption{The different coordinate frames of $AdS_{2}$. The global frame is the entire vertical strip (uncolored). The Poincar\'e patch is the largest triangular region. The black hole frame is a smaller triangular region and will always be colored in green.}
\label{fig:patches}
\end{figure}

 One of the primary motivations for studying the dilaton gravity theory is that it admits black hole solutions, which can be dynamically formed and deformed by throwing in matter from the boundary. In the $X^\pm$ coordinates, the static black hole background of mass $M=E$ is specified by the dilaton profile
 \bea
 \Phi^2 \! \is \! 1 \spc + \spc a \, \frac{1-\kappa \spc E\spc  X^+ X^-}{X^+ - X^-} \, ,
 \eea
 which manifestly solves the vacuum equations of motion with $T_{ab}=0$.
The coordinates $X^\pm$ are bounded to the region $\Phi^2>0$. Via the substitution
\bea
\label{BHx}
X^{+}(u)\! \! \is\!\! \frac{1}{\sqrt{\kappa E}}\tanh\Bigl(\sqrt{\ketw}\, u\Bigr)\,,\qquad \ \ 
X^{-}(v)  =  \frac{1}{\sqrt{\kappa E}}\tanh\Bigl(\sqrt{\ketw}\, v\Bigr) \, ,
\eea
we can write this black hole solition in a manifestly static form as
\bea
\label{BHmetr}
\dd s^{2} \is -\frac{4\kappa E}{\sinh^2\bigl(\sqrt{\ketw}\spc (u-v)\bigr)}\dd u\, \dd v\,, \\[2mm]
\Phi^2 \is 1 + a \sqrt{\kappa E} \coth\bigl(\sqrt{\ketw}\spc (u-v)\bigr) \, .
\label{BHdilaton}
\eea
The future and past horizon are found at $u = \infty$ and $v=-\infty$, which in terms of the $X$ coordinates translates to 
$X^+ =  1/{\sqrt{\kappa E}}$ and $X^- = - 1/{\sqrt{\kappa E}}$. In the limit $E\to 0$, one recovers the standard AdS${}_2$ metric (\ref{metricX}) in Poincar\'e coordinates, with dilaton profile $\Phi^2 = 1 + {a}/(X^+\!\! -\! X^-).$
As indicated in Figure \ref{fig:patches}, the black hole space-time outside the horizon can be embedded as a bounded triangular region within the Poincar\'e patch. Note that the size of the black hole patch shrinks with increasing $E$.

The black hole metric and dilaton in equations (\ref{BHmetr}) and (\ref{BHdilaton}) are periodic in imaginary time with period $
\beta = {\pi}/{\sqrt{\kappa E}}$. The Hawking temperature of the black hole is thus identified as
\bea
\label{temperature}
T = \frac 1 \pi {\sqrt{\kappa E}} \, .
\eea

\subsection{Black hole with shockwave}

\begin{figure}[H]
\begin{center}
\medskip
\includegraphics[width=0.26\textwidth]{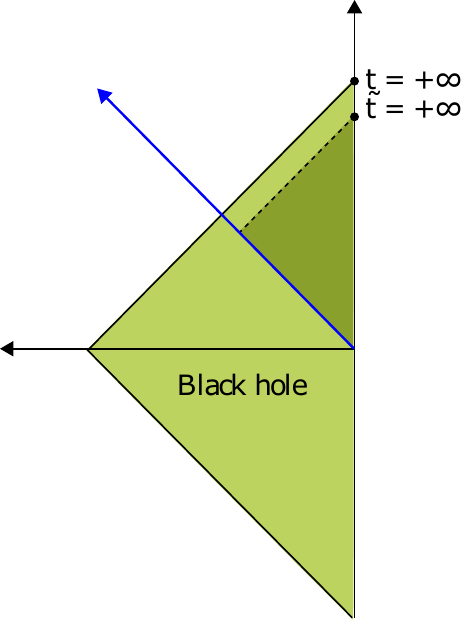}
\end{center}
\vspace{-2mm}
\caption{The black hole geometry with an infalling matter pulse. The pulse produces a new black hole space time with a slightly larger mass, indicated by the dark green region.
The point at future infinity gets shifted downwards.}
\label{fig:heavierbh}
\end{figure}

It is easy to generalize the black hole solution to include the effect of an infalling matter pulse. Suppose we start in a black hole space-time with mass $M=E$.
At time $t=t_1$, we send in  a massless particle with some small energy $
\delta E = \hbar \omega_1$.
The particle will fall via a light-like trajectory into the black hole, thereby increasing the black hole mass to
\bea
\tilde{E} \!\is\! E + \hbar \omega_1 \,.
\eea
The new black hole geometry has the exact same form as the old one, except with $E$ replaced by $\tilde{E}$ and with a new set of light-cone coordinates $(\tilde{u}, \tilde{v}$) related to the old coordinates via the matching condition
\bea
\frac{1}{\sqrt{\kappa E}}\tanh\Bigl(\sqrt{\ketw}\, (u-t_1)\Bigr) \! \is\! \frac{1}{\sqrt{\kappa \tilde{E}}}\tanh\Bigl(\sqrt{\kappa \tilde E}\, (\tilde{u}-t_1)\Bigr) \, .
\eea
As indicated in Figure \ref{fig:heavierbh}, the new black hole space-time of mass $\tilde{E}$ covers a smaller triangular region with the original black hole geometry. The shift $u-\tilde{u}$ that relates the old and new light-cone coordinate $u$ and $\tilde{u}$ is (for late times) equal to
\bea
\label{ushift}
u - \tilde{u} \! &\! \simeq \! &\!  \frac{\hbar \omega_1}{8 \sqrt{\kappa E^3}} \; e^{2\sqrt{\kappa E}\, (u-t_1)} \, .
\eea
This coordinate shift represents a gravitational shock wave interaction. Its exponential growth with time is a reflection of the exponential redshift near the horizon.

 Indeed, consider an outgoing signal that, without the infalling matter pulse, would have been destined to reach the boundary at some prescribed time $t_2$.
This outgoing signal follows the trajectory $u=t_2$. Due to the shock wave interaction with the infalling matter pulse, it instead reaches the boundary at a later time
$\tilde{u} = \tilde{t}_2$. From equation (\ref{ushift}) we read off that the signal is delayed by the exponentially growing amount 
\bea
\label{tshift}
\delta{t}_2\is  \frac{\hbar \omega_1}{8 \sqrt{\kappa E^3}} \; e^{2\sqrt{\kappa E}\, (t_2-t_1)} \, .
\eea
This is the familiar maximal Lyapunov behavior with $\lambda_L = 2\sqrt{\kappa E} = 2\pi T$ that characterizes the  dynamics between infalling and outgoing matter near a black hole horizon. In the next sections, we will rederive the formula (\ref{tshift}) from  the effective boundary theory.

\newpage

\section{Boundary dynamics}
\label{sect4}
It was argued in \cite{Almheiri:2014cka} that the time coordinate at the AdS${}_2$ boundary naturally becomes a dynamical variable, that interacts with the energy-momentum flow of the matter sector. We will denote the dynamical time by $\tau(t)$, where $\tau$ denotes a fixed reference time coordinate, which we choose to be the Poincar\'e time coordinate.
In other words, we will parametrize the boundary dynamics as a deformation relative to the vacuum solution
\bea
\label{vacsol}
\dd s^2 = -\frac{4\dd u \spc \dd v}{(u-v)^2}, \qquad \qquad \Phi^2 = 1 + \frac{a}{u-v} \, .
\eea
The preferred coordinate $t$ describes the time evolution along on the
\bea
\label{adsbd}
{\rm boundary \ of \ AdS_2} \is \bigl\{ \, u\spc =\spc v \spc  \equiv \spc t\, \bigr\} \, .
\eea
To derive the effective dynamics,  we adopt the view that an asymptotic observer is defined to follow a (very large but) constant value of the dilaton. 
The general background is specified in equations (\ref{metricX}), (\ref{Xpar}) and (\ref{dilat}). We  require that its boundary $u=v$ coincides with the unperturbed 
AdS${}_2$ boundary $X^+(u)= X^-(v)$. The dynamical boundary time is then defined via
\bea
X^+(t) \! \is \! X^-(t)  \, \equiv \, \tau(t) \, .
\eea

\subsection{Boundary equations of motion}

The metric and dilaton both diverge at the boundary. To extract the effective dynamics of $\tau(t)$, we introduce, following the customary procedure, an infinitesimally small regulator 
distance $\epsilon$ and move the boundary slightly inwards to  
\beq
u = t+\epsilon, \qquad v = t-\epsilon\, .
\eeq
The $X^\pm$ coordinates on the new boundary define two functions $(\tau(t),z(t))$
\bea
\frac 1 2\bigl( X^+(t+\epsilon) + X^-(t- \epsilon) \bigr) \is \tau(t) \, ,\\[1mm]
\frac 1 2 \bigl( X^+(t+\epsilon) - X^-(t- \epsilon) \bigr) \is \epsilon \spc  z(t) \, ,
\eea
related via
\bea
z(t) = \frac{\dd \tau(t)}{\dd t} \, .
\eea
The coordinate $\epsilon z(t)$ represents the distance between the dynamical boundary from the unperturbed AdS${}_2$ boundary. 

We now require that the dilaton has the same asymptotic form near the boundary as in the Poincaré patch. This requirement dictates that the dilaton takes a constant value at the dynamical boundary location 
\bea
\Phi^{2}(t) \! \is\!   \frac{a}{2 \epsilon\spc z(t)}\left[1 - \kappa \bigl( I_+(t) \spc + \spc I_-(t) \bigr) \right]\, = \frac{a}{2\epsilon} \, .
\, \ \label{eq:e105}
\eea
Hence we deduce that the dynamical boundary trajectory is described by
\bea
\label{zdynam}
{z(t)}\! \is\! 1 -  {\kappa} \bigl( \spc I_+(t) \spc +  I_-(t)\spc \bigr) \, , \\[3mm] \label{eq:e100}
I_+(t) \! \is \! \int^{\infty}_{\tau}\!\!\! \dd s\; (s - \tau)^{2}\, T_{++}(s) \, , \\[1.5mm]
I_-(t) \! \is\!  \int_{-\infty}^{\tau}\!\!\! \dd s\; (s - \tau)^{2}\, T_{--}(s) \, .
\eea
We can rewrite this equation in a somewhat more practical and suggestive form, by differentiating twice with respect the the dynamical time $\tau$.
Using that
\bea
\frac{\dd^2 I_\pm}{\dd\tau^2} \, = \, \pm 2 P_\pm ,  \quad && \qquad \ 
\frac{\dd }{\dd \tau}  \, \equiv \, \frac{1}{\partial_t\tau} \, \frac{ \dd  } {\dd t} \, ,\\[2.5mm]
P_+ \, = \,\int^{+\infty}_{\tau}\!\!\! \dd s\;  T_{++}(s), \ \ & & \ \ 
P_-  \, = \, - \int_{-\infty}^{\tau}\!\!\! \dd s\; T_{--}(s) \, ,
\label{ppm}
\eea
we deduce from (\ref{zdynam}) an equation of motion
\bea
\label{dynamo}
\frac 1 {2 \kappa }\, \frac{\dd^{2} z}{\dd \tau^{2}}\,  +\,  P_+\spc -\, P_- \is 0\,,
\eea
that relates the radial acceleration of the dynamical boundary to the difference between the left- and right-moving energy momentum flux. Equation (\ref{dynamo}) is reminiscent to the equation of motion that describes the recoil effect on a dynamical moving mirror due to Unruh radiation \cite{Chung:1993rf}. Differentiating once more gives
\bea
\label{jerk}
\frac 1 {2 \kappa }\, \frac{\dd^{3} z}{\dd \tau^{3}}\,  \is   T_{++} \spc -\, T_{--} \, ,
\eea
which has the characteristic form of the recoil due to radiation emitted by an accelerating particle. Note, however, that if we impose perfectly reflecting boundary conditions
 such that $T_{++} = T_{--}$, the force term on the right-hand side of equation (\ref{jerk}) vanishes. In this case the  left-hand side (the `jerk')  vanishes, and the boundary equation of motion reduces to the condition that the acceleration $\frac{\dd^{2} z}{\dd \tau^{2}}$ is constant in time.
 
The radial location $z(t)$ of the boundary is determined in terms of the time variable $\tau(t)$ via $z(t) = \frac{\dd \tau(t)}{\dd t}$. The equation of motion for the dynamical boundary 
time therefore reads
\bea
\label{boxed}
\quad \boxed{ \  
\frac 1 {2 \kappa} \, \frac{\dd^2}{\dd t^2} \log\Bigl( \frac{\dd \tau}{\dd t}\Bigr)\, +    \bigl(\spc P_+ -\spc P_-\spc \bigr) \, \frac{\dd \tau}{\dd t}{}_{\strut}^{\strut} \, = \,  0 \ } \,, \label{eomxt}
\eea
where $P_\pm$ are defined in equation (\ref{ppm}). The result (\ref{boxed}) is identical to the equation of motion derived by Almheiri and Polchinski \cite{Almheiri:2014cka}.  It is the starting point for our derivation of the effective Hamiltonian dynamics of the boundary time. 

\begin{figure}[H]
\centering
\includegraphics[scale=0.45]{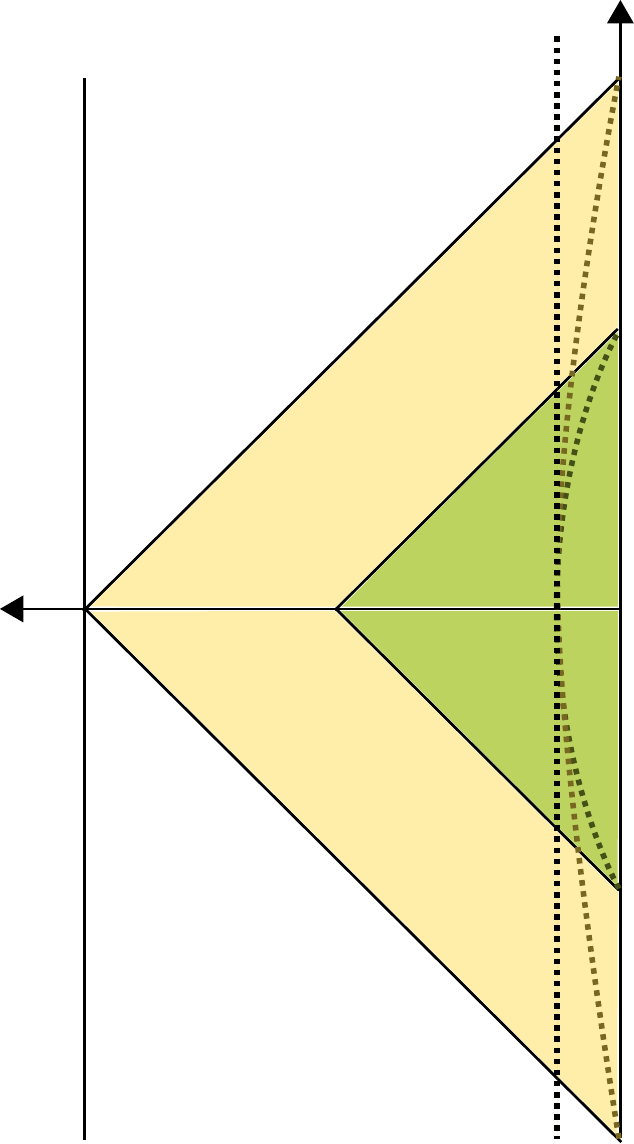}
\caption{Boundary trajectories in the different coordinate systems.}
\label{fig:boundary}
\end{figure}

Examples of boundary trajectories, in the same three frames as in Figure \ref{fig:patches}, are shown in Figure \ref{fig:boundary}.

\subsection{Boundary stress tensor}
\label{sect5}

Following the established holographic paradigm, one can extract a boundary stress tensor of the dual CFT via the standard holographic renormalization procedure.
The prescription involves varying the on-shell bulk action with respect to the boundary metric
\footnote{The boundary metric is obtained by removing the conformal prefactor from the bulk metric as: $ds^2 = \hat{\gamma}_{tt} d\hat{t}^2 = \frac{\gamma_{tt}}{z^2}dt^2$. } 
\bea
\bigl\langle\, \hat{T}_{tt}\spc \bigr\rangle = -\frac{2}{\sqrt{-\hat{\gamma}}}\frac{\delta S}{\delta \hat{\gamma}^{tt}} \, .
\eea
Since the boundary is one dimensional, there is in fact no momentum. The boundary stress tensor is identical to the boundary Hamiltonian 
\bea
H \is \hat{T}_{tt} \, .
\eea

The details of the calculation can be found in \cite{Almheiri:2014cka}. To get a self-consistent finite answer, the total bulk action needs to be supplemented by two boundary terms $S = S_{\rm bulk} + S_{\rm bd} + S_{ct}$ 
with $
S_{\rm bd} =   \frac{1}{8\pi G}\int \dd t \sqrt{-\gamma}  \Phi^2 K$ and $ S_{ct} \, =\, \frac{1}{8\pi G}  \int \dd t \sqrt{-\gamma} (1-\Phi^2)$
with  $K = \sqrt{g^{zz}}\frac{\partial_z\sqrt{-\gamma}}{\sqrt{-\gamma}}$ the extrinsic curvature. Utilizing standard arguments, one arrives at~\cite{Almheiri:2014cka}
\begin{equation}
\label{bdystress}
\bigl\langle\, \hat{T}_{tt}\spc \bigr\rangle =  \frac{\epsilon}{8\pi G} \bigl( e^{\omega} \partial_z\Phi^2 - e^{2\omega}(1-\Phi^2) \bigr) \, . 
\end{equation}
We would like to re-express this as a function of the dynamical boundary  time.  A straightforward calculation gives that $\Phi^2$ and $e^{\omega}$, given in equation (\ref{metricX}) and (\ref{dilat}), behave near the boundary as
\bea
e^{2\omega} \is\frac{1}{\epsilon^{2}} + \frac{2}{3}\{\tau,t\} + \mathcal{O}(\epsilon^{2})\,,\\[2mm]
\Phi^{2} \is 
\frac{a}{2\epsilon} + 1 -\frac{a}{3}\{\tau ,t\}\epsilon + \mathcal{O}(\epsilon^{2}) \, ,
\eea
\vspace{-1mm}
where 
\vspace{-1mm}
\bea
\bigl\{\smpc \tau,\smpc t\smpc \bigr\} = \frac  {{\tau}'''}{\tau'} - \frac 3 2\Bigl(\frac  {{\tau}''}{\tau'} \Bigr)^2 
\eea
denotes the Schwarzian derivative. Plugging this back into equation (\ref{bdystress}), we obtain
\bea
\label{tisderiv}
\boxed{\bigl\langle\, \hat{T}_{tt}\spc \bigr\rangle = -\frac 1 {2\kappa} \, 
\bigl\{\smpc \tau,\smpc t\smpc \bigr\} }\, .\label{bstress}
\eea
This appearance of the Schwarzian derivative is the salient feature of the AP model that underlies its likeness with the SYK model. 
It further seems to suggest a direct relationship with the anomalous transformation law of the energy momentum tensor in 2d CFT. Note, however, that the formula (\ref{tisderiv}) is derived
from classical bulk considerations, and that the boundary is one dimensional. So the relationship with 2d CFT seems mostly coincidental. For later reference, note that the 
Schwarzian derivative can be written as
\bea
\bigl\{\smpc \tau,\spc t\smpc \bigr\}=   -\frac 1 2( \partial_t \varphi)^2 + \partial^2_t \varphi, \qquad \varphi \equiv \log \partial_t \tau \, ,
\eea
which points to a possible relation with 1-dimensional Liouville theory.

Up to this point, everything has been set up referring to Poincar\'e coordinates as the most natural starting point. This is not necessary. In appendix \ref{pulseinglobal} we discuss the story so far when one uses global coordinates instead.

Some comments about the application of the results in this section to analyze pulse solutions are given in appendix \ref{multipulse2}.

\section{Hamiltonian formulation}
\label{sect6}

We have seen that the dynamical evolution of the AdS${}_2$ dilaton-gravity theory can be completely captured in terms of a dynamical boundary time variable $\tau(t)$, interacting with the energy momentum flux of the matter. This dimensional reduction is unsurprising: the dilaton and metric do not possess any local bulk degrees of freedom. In this section, we will
develop a practical Hamiltonian formulation of the boundary dynamics. The Hamiltonian language turns out to be convenient since it will automatically produce the canonical structure that can be used to compute the out of time-ordered commutators that can diagnose the Lyapunov behavior.\footnote{OTO commutators in 2d dilaton-gravity systems were first computed and studied in \cite{Chung:1993rf}\cite{Schoutens:1994st}.} 

As a warm-up, let us turn off the matter dynamics by setting $P_+ - P_- = \lambda$, with $\lambda$ some constant. The boundary equation (\ref{boxed}) then reduces to the Liouville-like 
equation 
\bea
\label{liouxv}
\quad \frac{1}{2\kappa} \partial_t^2 \varphi  + \lambda e^\varphi \is 0, \qquad e^\varphi = \partial_t \tau \, ,
\eea
which derives from the following action
\bea
\label{sliouville}
S \is   \int\! \dd t\, \Bigl(\smpc \frac{1}{4\kappa}(\partial_{t}\varphi)^{2} - \lambda e^{\varphi}  + \lambda\spc \partial_{t}\tau \Bigr) \, ,
\eea
where $\lambda$ is promoted to a dynamical variable, which via the $\tau$ equation of motion is set equal to a constant. The $\lambda$ equation of motion
imposes the identification $e^\varphi = \partial_t \tau$. Treating $\lambda$ as a Lagrange multiplier variable, we can choose to eliminate $\lambda$ and $\varphi$ via the constraint. The lagrangian (\ref{sliouville}) then reduces to the Schwarzian derivative 
\bea
\label{schwarz}
S \is -\frac 1 {2\kappa} \int \! \dd t \, \bigl\{\smpc \tau,\smpc t\smpc \bigr\} \,.
\eea
An alternative perspective is provided by integrating out $\tau$, which fixes $\lambda$ to a constant, and reduces this system to a 1d version of Liouville theory.

In principle we could proceed to describe the system using the reduced action (\ref{schwarz}). However, since the Schwarzian derivative action is higher order in time-derivatives, solving its equations of motion requires four integration constants, instead of the usual two. The associated phase space is therefore naturally four-dimensional instead of two-dimensional. We will therefore continue to work with the extended action (\ref{sliouville}). This will allow us to follow a more standard Hamiltonian treatment.

As a transition to a Hamiltonian language, we introduce the first order lagrangian
\bea
L \! \is \! \pi_\varphi\partial_t {\varphi} + \pi_\tau\partial_t{\tau} - H, \qquad \quad
H \, = \,  {\kappa}\spc {\pi_\varphi}^2 + e^{\varphi}\pi_\tau \, ,
\eea
\noindent
which leads to the canonical commutators $[\pi_\varphi,\varphi] = \frac{\hbar}{i}$ and $[\pi_\tau,\tau] = \frac{\hbar}{i}$ and to the
equations of motion
\bea
\label{momeqns}
\partial_t \varphi \! \! \is \! \! 2\kappa \pi_{\varphi} ,  \quad \partial_t \tau = e^\varphi,\quad  \partial_t\pi_{\varphi} =  -e^\varphi \pi_\tau \,, \quad  \partial_t\pi_{\tau} = 0 \, .
\eea
Setting $\pi_\tau = \lambda =$ constant reproduces (\ref{liouxv}).

\def\cft{{\mbox{\scriptsize \sc cft}}}

\def\lL{\mbox{\tiny L}}
\def\rR{\mbox{\tiny R}}

\subsection{Bulk matter}

To add the bulk matter system, we introduce an extra space-like coordinate $\sigma$, restricted to the positive half-line $\sigma \in [0,\infty)$.
Without coupling to the dynamical boundary, i.e. temporarily setting $\tau(t) = t$, the matter system evolves via the Hamiltonian $H=P_0$ given by the integral of the energy density 
\bea
P_0 \! \is\!   \int_{0}^{\infty}\!\!\! \dd \sigma\,T_{00}(\sigma) \, .
\eea
This Hamiltonian is a conserved quantity, provided we impose suitable reflecting boundary conditions at $\sigma=0$. For later reference,
let us also introduce the momentum operator, given by the integral of the momentum density
\bea
P_1 \is  \int_{0}^{\infty}\!\!\! \dd \sigma\,T_{01}(\sigma) \, .
\eea
The momentum  operator $P_1$ is {\it not} conserved, since the presence of the boundary breaks translation symmetry.
In the special case that the bulk matter system is described by a conformal field theory, we have
\bea
P_0 \is  \int_{0}^{\infty}\!\!\! \dd \sigma \bigl(T_{++}(t+\sigma)  +  T_{--}(t-\sigma)\bigr) \, \equiv \, P_+ - P_- \, ,\\[.5mm]
P_1 \is  \int_{0}^{\infty}\!\!\! \dd \sigma \bigl(T_{++}(t+\sigma)  -  T_{--}(t-\sigma)\bigr) \, \equiv \, P_+ + P_- \, ,
\eea
where $P_\pm$ denote the translation operators acting on the left- and right-moving sector of the CFT. The time derivative of both quantities is given by
\bea
\label{podot}
\partial_t{P}_0(t) \!\is \! -T_{++}(t) + T_{--}(t)\, =\, -{T}_{01}(t) \, ,\\[2.5mm]
\partial_t{P}_1(t) \!\is \! -T_{++}(t) - T_{--}(t)\, =\, -{T}_{00}(t) \, .
\eea
Energy is conserved as long as the momentum flux $T_{01}$ vanishes at the boundary.

Local CFT operators may be decomposed into a sum of chiral halves via
\bea
{\cal O}(\sigma, t) \is \sum_i {\cal O}^i_{\lL}(t+ \sigma)\spc {\cal O}^i_{ \rR}( t-\sigma) \, .
\eea 
These local bulk operators are in one-to-one correspondence with operators in the dual boundary system.
Before turning on the boundary dynamics, the dual operators are obtained by taking a suitable $\sigma \to 0$ limit of the bulk operators 
\bea
{\cal O}(t) = \lim_{\sigma \to 0} {\cal O}(\sigma, t) \, .
\eea  

To study the correlation functions of local boundary operators, or for other physical reasons, we can introduce a boundary interaction term via 
\bea
P_0 \to P_0 \, + \,  g(t) \, {\cal O}(t) \, ,
\eea
 with $g(t)$ some (small) time dependent coupling. The vacuum expectation value of the resulting time-evolution operator
is the generating functional of the boundary-to-boundary correlation functions
\bea
e^{-{\cal W}[g]}\!\! \is \Bigl\langle T\exp\Bigl( -i\!\! \int\!\! \dd t\, g(t) \spc {\cal O}(t) \Bigr)\Bigr\rangle \, ,\\[2mm]
{\cal W}[g] \is \int \!\! \dd t_1 \spc \dd t_2 \, \frac{g(t_1) \, g(t_2)}{(t_1 - t_2)^{2\Delta\!}} \; + \, \ldots \, .
\eea
Here the $\ldots$ indicate higher order interaction terms.

The above equations will all get modified due to the interaction of the CFT with the dynamical boundary. Indeed, in the 2d dilaton-gravity-matter theory,
setting $\tau(t) =t$ amounts to turning off the backreaction of the bulk dilaton and placing the matter CFT in the unperturbed AdS-background. The non-dynamical AdS${}_2$ geometry in Poincar\'e coordinates is conformally equivalent to a half plane, and the only geometric feature that the bulk CFT can interact with is the location of the boundary.

\subsection{Coupling the Boundary and the Bulk}

How do we couple the bulk CFT to the dynamical boundary system?
The non-interacting Hamiltonian $H_0$ of the combined bulk and boundary is given by
\bea
H _0 \is  \kappa \spc \pi_{\varphi}^{2} + e^{\varphi} \pi_\tau +  P_0 \, .
\eea
Let us write the dynamical boundary time as
\bea
\tau(t) = t + \epsilon(t) \, .
\eea
The deformation of the time variable is a geometric notion, so $\epsilon(t)$ can only act with the energy momentum tensor of the CFT at $\sigma=0$.
Since we are aiming to modify the reflecting boundary condition, it is necessary to allow the momentum flux $T_{01}(t)$ to be a non-trivial operator at $\sigma=0$.
Combining these two observations, we are led to consider the following interaction term 
\bea
\label{sint}
S_{\rm int} \is \int \! \dd t \, \epsilon(t) T_{01}(t) \nonumber \\[2mm]
 \is \int\! \dd t \, \dot{\epsilon}(t) P_{0}(t)  \, .
\eea
The first equation shows that the boundary interaction is local. To arrive at the second equation, we performed a partial integration, and used equation (\ref{podot}).
This second form of the interaction term turns out to be particularly convenient to incorporate a Hamiltonian formalism.

Based on the above discussion, we postulate that the coupling between the matter sector and the dynamical boundary theory is described via the total Hamiltonian 
\bea
\label{hami}
H \is  \kappa \spc \pi_{\varphi}^{2} + e^{\varphi}\bigl(\pi_\tau + P_0\bigr) \, ,
\eea
or written out more explicitly
\bea
\label{boxedH}
\quad \boxed{ H \, = \,  \kappa \spc \pi_{\varphi}^{2} + e^{\varphi} \pi_\tau + e^\varphi \int_{0}^{\infty}\!\!\! \dd \sigma\,T_{00}(\sigma)} \, .
\eea
As we will show, the equations of motion derived from this Hamiltonian exactly reproduce dynamical boundary equations of the dilaton-gravity-matter system described in the previous sections.  Since via the equation of motion, we can identify $\dot{\tau} = e^\phi$, the interaction term in the Hamiltonian (\ref{boxedH}) is equivalent to the interaction 
term in the action (\ref{sint}). Notice further that, although the coupling between $\varphi$ and the bulk matter system
extends throughout the bulk, the above discussion and equation (\ref{sint}) make clear that the above coupling between the CFT and the boundary system does in fact represent a local interaction.\footnote{The above construction of the Hamiltonian of the interacting bulk-boundary system in principle generalizes to non-conformal bulk systems. We expect, but have not shown, that the equivalence with the dilaton-gravity theory persists in this case. }

The Hamilton equations derived from (\ref{hami}) read
\bea
\partial_t \tau \nspc \is 
 e^\varphi, \qquad \qquad \ \partial_t \pi_\tau 
 \, = \, \spc  0\, ,\\[3mm]
\partial_t\varphi \nspc \is  
2 \kappa \, \pi_\varphi,\qquad \quad
\partial_t \pi_\varphi 
\, = \,  -e^{\varphi}\bigl(\pi_\tau +P_0 
\bigr)\, .
\eea
These combine to
\bea
\frac 1 {2\kappa} \partial_t^2 \varphi \! \is \!  \spc (\pi_\tau + P_0 )\spc  e^\varphi 
, \qquad \varphi \spc = \spc \log\bigl(\spc \partial_t \tau\bigr) \, ,
\eea
which (upon setting $\pi_\tau = 0$) coincides with the boundary equation of motion (\ref{boxed}).

As a further check on the correspondence, we note that the value of the Hamiltonian, when substituting the solution to the equations of motion, has the form:
\bea
\left.H\right|_{\mathrm{EOM}}  
\is   - \frac 1 {4\kappa} ( \partial_t \varphi)^2 + \frac 1 {2\kappa} \partial^2_t \varphi \, = \,  -\frac 1 {2\kappa} \left\{\tau,t\right\}\,,
\eea
which is the same as the expression for the boundary energy we obtained in the previous section. We have hence shown that the equations of motion of our Hamiltonian system indeed agree with those of the Almheiri-Polchinski model.

Local  operators in the bulk CFT evolve in time via $\partial_t{\cal O}  =   \frac i \hbar [\smpc H, {\cal O}\smpc] \, = \, \frac i \hbar \spc e^\varphi \spc  [\spc  P_0,\smpc {\cal O}\smpc].$ 
This integrates to
\bea
{\cal O}(\sigma, t) \is \sum_i {\cal O}^i_{\lL}(\tau(t)+\sigma)\spc {\cal O}^i_{ \rR}(\tau(t) - \sigma) \, . 
\eea
So in particular, we deduce that the local operators ${\cal O}(t)$ at the $\sigma\! =\! 0$ boundary become (operator valued) functions of the dynamical time coordinate $\tau(t)$.
\bea
\label{timo}
{\cal O}(t)\! \is\! \sum_i {\cal O}^i_{\lL}(\tau(t))\spc {\cal O}^i_{ \rR}(\tau(t)) \, \equiv \, \bar{\cal O}(\tau(t)) \, .
\eea

\subsection{$SL(2,\mathbb{R})$ symmetry}

The Schwarzian derivative $S(\tau,t) = \{ \tau, t\}$ is well known to be invariant under SL$(2,\mathbb{R})$, the group of M\"obius transformations $\tau \to \frac{a\tau+b}{c\tau+d}$ with $ad-bc=1$.
This invariance of the dynamical time variable plays a crucial role in the solution of the SYK model \cite{Kitaev}\cite{Polchinski:2016xgd}\cite{Maldacena:2016hyu}. As we will now show, the SL$(2,\mathbb{R})$ persists after coupling to the matter sector.

Consider the following set of generators
\bea
\label{lmin}
\quad {\ell}_{-1} \is \pi_\tau + P_0 \, ,
\\[2.5mm]
\label{lzero}
{\ell}_0 \is \tau\left(\pi_\tau + P_0\right) + \pi_\varphi  \, ,
\\[1mm]
\label{lplus}
{\ell}_1 \is  \tau^2(\pi_\tau + P_0) + 2 \tau \pi_\varphi - \frac 1 \kappa e^{\varphi} \,.
\eea
Using the canonical commutation relations, it is easy to verify that these charges commute with the total Hamiltonian $H$ (\ref{hami}) and satisfy an $\mathfrak{sl}(2,\mathbb{R})$ algebra:
\bea 
\left[\ell_0,\ell_{\pm 1}\right] \! \is \!  \mp i\hbar \ell_{\pm 1}\, ,\nonumber\\[-2mm] \\[-2mm]
\left[\ell_1, \ell_{-1}\right] \! \is \!  2 i\hbar \ell_0 \, .\nonumber
\eea
These charges are not completely independent from the Hamiltonian: $H$ is equal to the quadratic Casimir  
\bea
\quad \boxed{\, \frac 1 2 \{\ell_1, \ell_{-1} \} -  \ell_0^2 \, =\, - \frac 1 \kappa H\,} \, .
\eea

To analyze the SL$(2,\mathbb{R})$ symmetry a bit further, we temporarily again decouple the matter sector and set $P_0 =0$.
The three charges  then generate the following infinitesimal transformations
\bea
\ell_{-1}&: & \ \delta \tau =  \epsilon \, ,\nonumber \\[1.5mm]
\ell_0 &: & \ \delta \tau =  \epsilon \tau \quad \delta \varphi = \epsilon, \quad \delta \pi_\tau =  - \epsilon \pi_\tau \, , \\[1mm]
\ell_1&: & \ \delta \tau = {\epsilon}\tau^2, \quad \delta \varphi = 2\tau\epsilon, \quad \delta \pi_\varphi =  \frac \epsilon \kappa e^{\varphi}, \quad \delta \pi_\tau =  - 2\epsilon \tau \pi_\tau - 2 \epsilon \pi_\varphi \,, \nonumber
\eea
which exponentiate to
\bea
\label{sltwoa}
 \tau \! &\! \to\!  & \! \frac {a\tau+b}{c\tau+d} ,\qquad \qquad \quad  \;
\pi_\tau\, \to \, (c\tau +d)^2 \tilde\pi_\tau  + 2c (c\tau + d)\tilde\pi_\varphi \, ,\\[1mm]
e^\varphi\! &\!  \to\!  & \!   \frac{e^\varphi}{(c\tau +d)^2},
 \qquad\qquad 
\pi_\varphi\, \to\, \tilde\pi_\varphi  \, ,
\label{sltwob}
\eea
with\\[-7mm]
\bea
\label{extra}
\tilde\pi_x \! &\!\! = \!\! & \!   \pi_x - \partial_x\alpha \qquad \qquad  \quad \ \alpha \, =\, \frac{c}{\kappa(c\tau + d)}\, e^\varphi \, .\qquad\qquad\qquad
\eea 
Equations (\ref{sltwoa})-(\ref{extra}) define a family of canonical transformations on phase space,  that leave the Hamiltonian invariant.

What is the physical role of the SL$(2,\mathbb{R})$ symmetry in the bulk dilaton-gravity theory? The AdS${}_2$ space-time is a symmetric space, on which M\"obius transformations act as  isometries. However, this symmetry is broken via the condition that the dilaton background $\Phi^2$ is required to take the asymptotic form (\ref{vacsol}) with non-zero constant $a$. Intuitively, one can view the dynamical time variable $\tau$ as a pseudo-Goldstone mode, that arises from the `spontaneous
symmetry breaking' of the AdS${}_2$ isometry group, on which the SL$(2,\mathbb{R})$ symmetry is non-linearly realized \cite{Kitaev}\cite{Polchinski:2016xgd}\cite{Maldacena:2016hyu} \cite{Maldacena:2016upp}	({\it c.f.} \cite{Jackson:2014nla}). Alternatively, we can `gauge fix'  the SL$(2,\mathbb{R})$ symmetry
by prescribing the asymptotic behavior of $\tau(t)$ by setting $\tau(t) = t$  in the far past and far future. At intermediate times, however, the M\"obius symmetry group then re-emerges as a quasi-static non-linearly realized symmetry.

\subsection{Virasoro symmetry}

The SL$(2,\mathbb{R})$ symmetry generators (\ref{lmin})-(\ref{lplus}) also act on the matter sector, via the commutator with the matter Hamiltonian $P_0$.
A useful physical way to characterize the SL$(2,\mathbb{R})$ action is by pointing out that the following general class of operators
\bea
A_{\lL}(\spc f, t\spc ) \is \int_0^\infty \! \dd \sigma \, f(\sigma + \tau(t)) \, A_{\lL}(\sigma,t)\, ,\\[2mm]
A_{\rR}(\spc f,t\spc ) \is \int_0^\infty \! \dd \sigma \, f(\sigma - \tau(t)) \, A_{\rR}(\sigma,t)\, ,
\eea
where $A_{\lL,\rR}$ denote  chiral CFT operator and $f(\sigma)$ denotes some arbitrary test function, are SL$(2,\mathbb{R})$ invariant -- up to boundary terms at $\sigma = 0$,
which can be avoided by choosing the support of the test function appropriately. So in particular, this general class of operators commute with the Hamiltonian, and thus represent conserved quantities. 

A particularly important class of conserved quantities are the frequency eigen modes of the energy momentum tensor
\bea
P_\omega \is 
 \int_0^\infty\!\!\! \dd \sigma\, \Bigl( \spc e^{i\omega  ( \tau(t) + \sigma)}\, T_{++}(\sigma,t) + e^{i\omega  (\tau(t)-\sigma)}T_{--}(\sigma,t)\spc \Bigr) \,.\quad
\eea
Their conservation is predicated on the reflection condition 
\bea
T_{++}(t) = T_{--}(t)
\eea
at the $\sigma = 0$ boundary. The charges $P_\omega$ satisfy the Virasoro algebra with a central extension
\bea
[ P_\omega, P_{\omega'}] = (\omega-\omega') P_{\omega + \omega'} + \spc \frac{\pi}{6} c \, \spc \omega^3  \spc \delta(\omega-\omega') \,.
\eea
These charges all commute with the SL$(2,\mathbb{R})$ generators (\ref{lmin})-(\ref{lplus}): the Virasoro symmetry should indeed not be seen as an extension of the M\"obius group described in the previous subsection. The main lesson here is that the interaction between the CFT and the dynamical boundary preserves half of the infinite dimensional conformal symmetry.
In this respect, the boundary dynamics defines a proper boundary CFT. Note, however, that our discussion so far has been semi-classical. Substantial more work is needed to establish whether the model described here defines a fully consistent theory.

\section{Commutators and chaos}
\label{commutatorsection}
The boundary dynamical system is integrable, and the equations of motion for $\tau$ and $\varphi$ can be explicitly integrated. In this situation,
the canonical formalism is particularly useful for computing commutators between operators $W(t_2)$ and $V(t_1)$ defined at different times. The strategy is simple: one 
writes the general solution to the equation of motion, and expresses, say, the operator $W(t_2)$ in terms of phase space quantities defined at the earlier time $t_1$. The 
commutator $[W(t_2),V(t_1)]$ is then easily computed from the canonical commutation relations.

In this section we will use the above strategy to show that the expectation value of the commutator squared, evaluated at finite energy $E$ (or temperature $T = \frac 1 \pi \sqrt{\kappa E}$) and over an intermediate range of time intervals $t_2-t_1$ =  $t'_2-t'_1$ , behaves as follows
\bea
\label{oto}
\frac{\bigl\langle \bigl[ \smpc V(t_1) , W(t_2)\smpc \bigr]  \bigl[\smpc V(t'_1) , W(t'_2)\smpc \bigr]\bigr\rangle_{}}
{\langle V(t_1)V(t_1')\rangle^{}\langle W(t_2) W(t_2')\rangle^{}}  \! & \! \simeq\! & \!  (\omega_2 \delta t_2)^2 \, ,\spc
\eea
with ({\it c.f.} equation (\ref{tshift}))
\bea
\label{newtshift}
\delta t_2 \, = \,\frac{\hbar \omega_1}{16 \sqrt{\kappa E^3}}  e^{2 \sqrt{\kappa E}\spc (t_2-t_1)} \, ,
\eea
where $\hbar \omega_1$ and $\hbar \omega_2$ denote the energy injected or absorbed by the respective operators $V$ and $W$.\footnote{Evidently, we could equally well have chosen to write the r.h.s. of (\ref{oto}) as $(\omega_1\delta t_1)^2$.} The calculation is based on the formula (\ref{timo}), which allows us to express the time evolved CFT operators  $V(t_1)$ and $W(t_2)$ in the presence of the dynamical boundary in terms of the unperturbed CFT operators $\bar{V}(t_1)$ and $\bar{W}(t_2)$ via 
\bea
V(t_1)\! \is\!  \bar{V}(\tau(t_1)),  \ \qquad W(t_2)\, = \, \bar{W}(\tau(t_2)) \, .
\eea
So, ignoring the part of the commutator that follows from the unperturbed CFT dynamics, we deduce that the relevant commutator is given by
\bea
\label{vwcomtt}
\bigl[\smpc V(t_1) , W(t_2)\smpc  \bigr]\! & \! \simeq\! & \! [\tau(t_1) , \tau(t_2)] \;  \frac{\partial_t V(t_1)}{\partial_t\tau(t_1)}\, \frac{\partial_t W(t_2)}{ \partial_t \tau(t_2)} \, .
\eea
Thanks to this relation, our task is simplified to computing the commutator $[\tau(t_1) , \tau(t_2)]$ between the dynamical time variable evaluated at different times $t_1$ and $t_2$.
The Lyapunov growth of the commutator in (\ref{oto}) saturates the chaos bound $\lambda_L = 2\pi T$.

\subsection{Classical trajectory}

The general classical boundary trajectory with energy $H=E$ is given by
\bea 
\label{solphi}
 e^{-{\varphi(t)}} \is  \pze
 \, {\cosh^2\Bigl(\sqrt{\ketw} \spc (t-t_0) \Bigr)}\,,\label{phisol} \\[3mm]
\pi_{\varphi}(t) \is-  \sqrt{\twek} \tanh\Bigl(\sqrt{\ketw}\spc (t-t_0) \Bigl)\, , \label{pisol} \\[3mm]
 \tau(t) -\tau(t_0) 
\is
\oops \, \sqrt{\twek} 
\, \tanh\Bigl(\sqrt{\ketw}\spc (t-t_0) \Bigr) 
\, \,.\label{tausol}
\eea
Using the hyperbolic identity $\tanh^{2} x + \cosh^{-2} x = 1$, we immediately verify that
\bea
E \is \kappa  \pi_{\varphi}^{2} + e^{\varphi}(\pi_\tau+ P_0)\,.
\eea
All quantities on the right-hand side of equations (\ref{solphi})-({\ref{tausol}) (except for $t$) are integration constants. We have expressed them in terms of the conserved quantities $E$, $\pi_\tau$ and $P_0$, since this will allow us to keep track of commutation relations. Indeed, we can view these equations as operator identities, that in particular relate the dynamical time variable $\tau(t)$ at time $t$ to the time variable $\tau(t_0)$ at $t=t_0$. Note that the dynamical time $\tau(t)$ runs over a finite range. In other words, at late and early time, the $\tau$ evolution is slowed down by an infinite redshift factor. Indeed, $\tau(t)$ has the natural physical interpretation of the time evolution as seen by an infalling observer.

The trajectory (\ref{solphi})-(\ref{tausol}) reproduces the classical solution (\ref{BHx})-{\ref{BHdilaton}) for the black hole geometry of mass $M=E$, provided we set
\bea
\label{replace}
\pi_\tau + P_0 = E \, .
\eea
Below, we will use this equality to simplify the final expression after we have computed the commutator between $\tau(t_2)$ and $\tau(t_1)$.

\subsection{Computation of $[\tau(t_1),\tau(t_2)]$}

We choose two time instances $t_1$ and $t_2$, one early and one late
\bea
t_1 \spc < \spc t_0 \spc < \spc  t_2 \, ,
\eea
such that the time differences $t_2-t_0$ and $t_0-t_1$ are both large compared to the characteristic time scale $1/T \sim 1/\sqrt{\kappa E}$. In this regime, we can use the following trick
\bea 
[\tau(t_1), \tau(t_2)] \is \frac 1 2 \bigl[\smpc \tau(t_1)+ \tau(t_2), \tau(t_2)- \tau(t_1)\smpc \bigr]\, ,
\eea
where (using equation (\ref{tausol}) and that  $\tanh x \to \pm 1$ for large positive or negative~$x$) the sum and difference are equal to 
\bea
\tau(t_1)+ \tau(t_2) \, \simeq \, 2\tau(t_0) \qquad \ \ \tau(t_2) -\tau(t_1) \!& \! \simeq  \! & \frac 2 {\pi_\tau+ P_0}\, \sqrt{\frac{E}{\kappa }} \, ,
\eea
up to exponentially small corrections. We now easily compute
\bea
[\pi_\tau,\tau(t_0)] = -i \hbar, \qquad [ E, \tau(t_0) ] \is -i\hbar  \partial_t \tau(t_0) \, =\, \frac{ i \hbar \spc E}{\pi_\tau + P_0} \, ,
\eea
so that (using $[\tau(t_0),P_0] = 0$)
\bea
\label{taucomm}
\boxed{\ [\tau(t_1) , \tau(t_2)] \, = -  \spc \frac {\,i \hbar} {(\pi_\tau+P_0)^2}\, \sqrt{\frac{E}{\kappa }}\ } \, .
\eea

\subsection{Exchange algebra}

We can now put everything together. We adopt the choice (\ref{replace}) at this moment. 
At late and early times, the $\tau$ evolution is slowed down by an exponential
redshift factor 
\bea
\label{redshift}
\qquad\qquad \partial_t \tau(t) \! \! & \! \simeq \! &  \!  4 \, e^{\mp 2 \sqrt{\kappa E}\,  (t- t_0)}\qquad \quad  \mbox{\small $t \to \pm \infty$} \, .
\eea
Combining equations (\ref{vwcomtt}), (\ref{taucomm}) and (\ref{redshift}) we arrive at the final result for the commutator between time separated local operators\footnote{This commutation relation is very similar to the one derived for 2d dilaton-gravity with zero cosmological constant in \cite{Chung:1993rf}\cite{Schoutens:1994st}. }
\bea
\bigl[\smpc V(t_1) , W(t_2)\smpc \bigr] \, = \, i\hbar \, \frac{\spc e^{2\sqrt{{\kappa E}}\spc (t_2-t_1)}\,}{16 \sqrt {\kappa E^3}}  \, \partial_t V(t_1)\,  \partial_{t}\smpc W(t_2) \,.
\eea
The physical interpretation  of this result becomes more evident if we assume that the operators $V(t_1)$ and $W(t_2)$ inject and absorb some given amount of energy equal to $\hbar\omega_1$ and $\hbar\omega_2$, respectively,  so that we can replace 
 $i\partial_t V(t_1) = \omega_1 V(t_1)$ and  $i \partial_t W(t_2) = \omega_2 W(t_2)$. We will still assume that each operator remains localized in time. 
 The above commutation relation can then be rewritten as an exchange algebra
\bea
\label{exchange} 
\boxed{\ V^{{}_{\strut}}(t_1)\, W{{}_{{}^{\strut}}}(t_2) \, = \, W(t_2+\delta t_2) \, V(t_1-\delta t_1)\ } \,,
\eea
with $\delta t_2$ the time shift given in (\ref{newtshift}) and $\omega_1 \delta t_1 = \omega_2 \delta t_2$.  This exchange algebra expresses the physical effect of the gravitational shockwave, caused by the infalling
perturbation created by $V(t_1)$, on the outgoing trajectory of the signal detected by $W(t_2)$. Similarly, the time shift $\delta t_1$ is a recoil effect on the incoming perturbation.
Note, however, that our result here for the time delay $\delta t_2$ differs by a factor of 2
from the result (\ref{tshift}) found by the classical calculation using a delta function matter pulse. We suspect that the classical computation gives an 
overestimate of the coordinate shift  as it ignores the recoil effect on the infalling wave. 

The exchange algebra (\ref{exchange}) implies the exponential behavior (\ref{oto})-(\ref{newtshift}) of the OTO four-point correlation function. However, it gives more detailed physical information: it exhibits that the dynamical influence of the early perturbation has a purely geometric effect (in the form of an exponential time delay) on the outgoing signal.

\section{Quantum effects}
\label{sect3}

In this section we discuss some quantum aspects of the model and investigate some consequences of the conformal anomaly of the bulk CFT. 
The conformal anomaly shows up as an inhomogeneous transformation property of the chiral energy momentum tensor,
or as a trace anomaly of the covariant energy momentum tensor
\bea
\label{trace}
T_{uv}\! = \! -\frac{c}{12\pi}\partial_u \partial_v \omega \, ,
\eea
So it is natural to ask the question:
How does this quantum correction show up in the effective action for the dynamical boundary theory?

\subsection{Boundary interaction}

A practical way to study correlation functions of the boundary theory is to add interaction terms to the Hamiltonian, say, of the form
\bea
H \is e^{\varphi(t)} P_0  +
g(t) {\cal O}(t) + \xi(t) T_{00}(t) \, ,
\eea
with ${\cal O}(t)$ some local CFT primary operator of conformal dimension $\Delta$ and $T_{00}(t)$ is the time component of the CFT energy-momentum tensor.
We split the interacting Hamiltonian as a sum $H  = P_0 + H_{\rm int}$ 
of an unperturbed CFT Hamiltonian  and an interaction term $
H_{\rm int} = g(t) {\cal O}(t) +  \epsilon(t) T_{01}(t) + \xi(t) T_{00}(t)$
where $\epsilon(t) = \tau(t)-t$.
We can then define an effective action via
\bea
e^{-{\cal W}[\tau,\xi, g]}\!\! \is \Bigl\langle Te^{\mbox{\footnotesize $-i\! \int\! \dd t\, H_{\rm int}$}} 
\Bigr\rangle_{\! \cft} \, ,
\eea
where the expectation value is taken in the unperturbed CFT. 

The effective action ${\cal W}[\tau,\xi,g]$ is a non-local functional of the couplings $(\tau(t),\xi(t), g(t))$. It plays a dual role:
via its $g(t)$ and $\xi(t)$ dependence, it provides a generating functional for boundary-to-boundary correlation functions of the ${\cal O}(t)$ and $T_{00}(t)$.
Via its $\tau(t)$ dependence, it supplies an extra term in the Schwarzian derivative boundary action, generated by the presence of the classical couplings $\xi(t)$ and $g(t)$.
Up to quadratic order in the couplings, we have
\bea
\label{wone}
{\cal W}[\tau,g,\xi]\! \is \!\! \int \!\! \dd t_1  \dd t_2\nspc \left(g(t_1) \spc g(t_2) \left(\frac{\dot\tau(t_1) \dot \tau(t_2)}{(\tau_1 - \tau_2)^2}\right)\raisebox{6pt}{${\!}^\Delta$}\!\! +  c \, \spc \xi(t_1) \spc \xi(t_2) \left(\frac{\dot\tau(t_1) \dot \tau(t_2)}{(\tau_1 - \tau_2)^2}\right)\raisebox{6pt}{${\!\nspc}^2$} \nspc + ...\, \right) \, . \ \  \nonumber\\[-2mm]
\eea

\subsection{Non-linear effective action}

We now set $g(t)=0$. The effective action ${\cal W}(\tau, \xi)$ can then be computed to all orders as follows.
Turning on the coupling $\xi(t)$ amounts to a general deformation of the boundary trajectory, by allowing it to move away from $z=0$. The boundary location
is then specified by two coordinates 
\bea
\label{tzboundary}
X^\pm(t) = \tau(t) \pm \sigma(t) \, ,
\eea
where $\sigma(t)$ is related to $\xi(t)$ via $\sigma(t) = \xi(t) \dot \tau(t)$ (for infinitesimal $\xi(t)$).
We may think of $(\tau(t),\sigma(t))$ as specifying the trajectory of a moving reflecting mirror at the end of space. In the correspondence with the Almheiri-Polchinski  model,  $\tau(t)$ is a dynamical variable and $\sigma(t)$ is held fixed at $\sigma=0$. We are free, however, to introduce the transverse location $\sigma(t)$ as a non-dynamical variable, that can be used as
a source that couples to $T_{00}(t)$. By varying $\sigma(t)$ we are able to inject energy and momentum into the system.

The time evolution in the presence of a given boundary trajectory $\bigl(\tau(t),\sigma(t)\bigr)$ is described by the time-dependent Hamiltonian 
\bea
\label{defh}
 H_{\cft} \bigl(\tau(t), \sigma(t))\! \is\!\nspc  \int_0^\infty \!\!\!\nspc \dd\sigma \spc\bigl( \dot\tau(t) T_{00} + \dot{\sigma}(t) T_{01}\bigr)\,=\,  \dot\tau(t) P_0 + \dot \sigma(t) P_1 \, .
\eea
As before, we split $H  = P_0 + H_{\rm int}$ and define an effective action ${\cal W}[\tau,\sigma]$ by taking the expectation value of the corresponding time evolution operator.

We wish to obtain an exact expression for ${\cal W}[\tau,\sigma]$.  We choose to simplify our task by rotating to Euclidean signature, which eliminates most of the subtleties associated with the choice of vacuum boundary conditions.\footnote{The real-time version of this problem was recently studied in \cite{Pimentel:2015iiv}.}
Based on the result of earlier investigations \cite{Pimentel:2015iiv}\cite{Chung:1993rf}} of a very similar problem, we propose that the expression for the effective action ${\cal W}[\tau,\sigma]$ can be cast in the form
\bea
\label{effact}
{\cal W}[\tau, \sigma] \is \frac{c}{96\pi}\int \!\! \dd t_1\spc   \dd t_2   \,  \dot\rho(t_1) \smpc \dot\rho(t_2)\, G_{[\tau,\sigma]}(t_1, t_2) \, ,\\[2mm]
 & & \quad \rho =
 \log\Bigl(\frac{\dot{\tau} +i  \dot \sigma}{\dot\tau -  i\dot \sigma}\Bigr),       
\eea 
where $ G_{[\tau,\sigma]}\bigl(t_1, t_2\bigr)$ denotes the scalar Green function, defined
on the half-plane with (\ref{tzboundary}) as its boundary, between the two boundary points $(\tau(t_1),\sigma(t_1))$ and $(\tau(t_2),\sigma(t_2))$.
 Finding an explicit expression for the Green function is hard, as it involves solving a highly non-trivial Riemann-Hilbert type problem \cite{Pimentel:2015iiv}.
Note that the right-hand side vanishes if we set $\dot{\sigma}=0$. It is easy to see that the second order expansion in $\dot{\sigma}$ reproduces the second term in (\ref{wone}), where a $4\pi^2$ factor coming from the different normalization conventions of the stress tensor is left implicit.
We will now motivate the formula (\ref{effact}). For simplicity, we will work in the regime $c\gg 1$, though we expect (\ref{effact}) to hold in general.

\def\zb{{\tilde x}}

The Hamiltonian (\ref{defh}) can be viewed as describing a CFT propagating on a {\it flat} space-time with metric $\dd s^2 = \partial X \bar\partial \bar{X} \dd x \dd\bar{x}$ and a boundary at $x=\bar{x} = t$.  
The metric dependence of a CFT partition function is uniquely prescribed by the conformal anomaly and  Ward identities. The trace anomaly (\ref{trace}) and the extra terms in (\ref{general}) and (\ref{general2}) are accounted for via the non-local Polyakov action 
\bea
\label{polact}
S_P = - \frac{c}{96\pi} \int\! \dd^2x\spc \sqrt{g}\, R\,  \square^{-1} R \, .
\eea 
$S_P$ can be recast into a local form by introducing an auxiliary scalar field $\chi$ with action
\bea
\label{chiact}
S_{\chi} \is -\frac{c}{24\pi}\int\! \dd^{2}x\spc \sqrt{g}\left[\partial_\mu \chi \partial^\mu \chi + \chi R\right] - \frac{c}{12\pi}\int\! \dd t\spc \sqrt{\gamma}\chi K \, ,
\eea
where $K$ denotes the extrinsic (i.e. geodesic) curvature of the boundary trajectory. 
Integrating out $\chi$ yields back the Polyakov action.\footnote{The prefactor $\frac{c}{24\pi}$ in (\ref{chiact}) should in fact be replaced by $\frac{c-1}{24\pi}$. This subtlety is sub-leading at large $c$, and does not affect the final conclusion.} The extra boundary term is needed to reproduce the correct form of the Polyakov action for a space-time with a boundary.\footnote{In fact, the above action (\ref{polact}) itself should be augmented by boundary terms as well.}

In the conformal gauge $\dd s^2 = e^{2\omega(x,\bar{x})} \dd x \dd \bar{x}$ the $\chi$ action simplifies to  
\bea
\label{chiact2}
\quad S_{\chi} \is -\frac{c}{6\pi}\int\! \dd^{2}x\spc  \partial \chi \bar{\partial} \chi - \frac{c}{12\pi}\int\! \dd t \spc \chi \spc (\partial \omega - \bar{\partial} \omega)_{\strut|x=\bar{x}} \, ,
\eea
where we used that $R=0$.
Since $e^{2\omega} =\partial X \bar\partial \bar X$ we have at the boundary
\bea
\partial\omega - \bar\partial \omega = \frac 1 2\bigl( \partial\log \partial X - \bar\partial\log\bar\partial \bar{X}\bigr) =  \frac 1 2 \partial_t  \log\Bigl(\frac{\dot{\tau} +  i\dot \sigma}{\dot\tau - i\dot \sigma}\Bigr) \, .
\eea 
Plugging this into (\ref{chiact}) and performing the $\chi$ integral gives the announced result (\ref{effact}).

\subsection{Black hole evaporation}
\label{evapobh}

The conformal anomaly is directly connected with the appearance of Hawking radiation.
Energy-momentum conservation requires that the light-like components $T_{uu}$  and $T_{vv}$ receive an extra contribution
\bea
\label{general}
T_{uu} \! \is \! -\frac{c}{12\pi}\bigl( (\partial_{u}\omega)^2 + \partial_{u}^{2}\omega\bigr) \; + :T_{uu}: \, ,\\[2mm]
\label{general2}
T_{vv} \! \is \! -\frac{c}{12\pi}\bigl( (\partial_{v}\omega)^2 + \partial_{v}^{2}\omega\bigr) \; + :T_{vv}: \, ,
\eea
where the second term is the chirally conserved, but non-covariant, normal-ordered energy momentum tensor:
neither term is separately covariant but the sum is.  
Both of these have merit on their own. The covariant tensor is the one that should be inserted into Einstein's equations and hence is responsible for backreaction. The normal ordered stress ``tensor'' corresponds to the stress tensor that would be measured by local observers using detectors calibrated to their vacuum. 

The bare AP model without any boundary perturbation leads to perfect reflection at the AdS boundary. So an AdS${}_2$ black hole will fill its surroundings with an eternal heat bath of virtual particles. We can simulate an evaporation process by allowing particles to escape from the thermal atmosphere, and thus relaxing the condition that $T_{01} = T_{uu} - T_{vv}=0$ at the boundary. From the bulk side, a straightforward computation gives
that the time derivative of the ADM Hamiltonian (\ref{tisderiv}) is merely the net flux of energy thrown into the spacetime\footnote{
The same result is easily derived from the Hamiltonian formulation presented in section \ref{sect6}. Assuming that the source of explicit time dependence comes from the pure matter
Hamiltonian $P_0$, and using that $\partial_t{P_0} = (T_{--} - T_{++})\partial_t \tau$ we deduce that
$$
\frac{\dd H}{\dd t} =  e^\varphi\, \frac{\partial P_0}{\partial t} \, = \, \bigl(T_{--} - T_{++}\bigr)\Bigl(\frac{\partial\tau}{\partial t}\Bigr)^2 \, ,
$$
which is equation (\ref{energyrate}).}
\begin{equation}
\label{energyrate}
\boxed{\frac{\dd E}{\dd t} = -\frac{1}{2\kappa} \frac{\dd }{\dd t} \left\{\tau,t\right\} = T_{vv}(t) - T_{uu}(t)}\, ,
\end{equation}

For all of the cases of interest to us, the conformal anomaly cancels in this expression: $T_{vv}(t) - T_{uu}(t) = \, :T_{vv}(t): - :T_{uu}(t):\, $.
We can now use this equation to compute the energy loss of a black hole due to the Hawking evaporation. Suppose we start from the vacuum AdS${}_2$ space-time in Poincar\'e coordinates. We create a black hole at some time $t=0$, by sending in a matter pulse with total energy $E_0$ (Figure \ref{fig:blackhole}).
Assuming that the initial state did not contain any outgoing matter,
 we learn that
 \bea
 \label{incoming}
T_{++} \is   :T_{++}:  \,\,\, = \,  0 \, .
 \eea

\begin{figure}[h]
\begin{center}
\medskip
\includegraphics[width=0.36\textwidth]{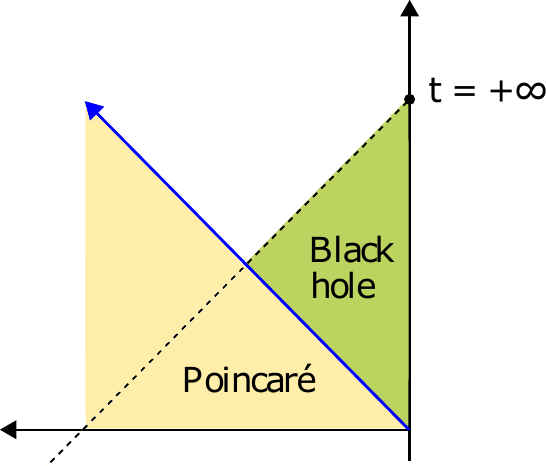}
\end{center}
\vspace{-2mm}
\caption{Creation of a black hole by sending in a pulse in the Poincar\'e patch. The dashed line represents the black hole horizon as described in the black hole frame.}
\label{fig:blackhole}
\end{figure}

Next we imagine placing a perfect detector that absorbs 
every physical particle that reaches the $z=0$ boundary. In terms of the energy momentum tensor, this amounts to imposing perfect absorption boundary condition.
In equations, this means that there is no energy-momentum flux leaving the boundary into the AdS${}_2$ space time:
\bea
:T_{vv}(t):  \is  0 \,. 
\eea
\def\ccc{{\raisebox{0pt}{$c$}}}
The outgoing energy momentum tensor, on the other hand, we know is non-zero 
\bea
:T_{uu}(u): \is  (\partial_u X^+)^2:T_{++}: + \frac{\ccc}{24\pi}\left\{X^+,u\right\} \, ,
\eea
which using (\ref{incoming}) tells us that on the boundary at $t>0$
\bea
:T_{uu}(t): \is  \frac{\ccc}{24\pi}\left\{\tau,t\right\} \, .
\eea

Hence we arrive at the somewhat tantalizing result that at times greater than $t=0$, the rate of change of the energy is proportional to the energy 
\begin{equation}
-\frac{1}{2\kappa} \frac{\dd\ }{\dd t} \left\{\tau,t\right\} = \frac{\ccc}{24\pi}\left\{\tau,t\right\} \, .
\end{equation}
So the energy decays exponentially 
\begin{equation}
{E(t) = E(t_0) \exp\left(- A (t-t_0) \right)} \, , 
\end{equation}
with evaporation rate 
\bea
A \is \frac{\kappa \spc c}{12\pi} \, = \, \frac{2G\spc c}{3a} \, . 
\eea
So the characteristic timescale associated to this evaporation process is $t_{\rm evap} =  1/A$.\footnote{An exponentially decaying profile could have been anticipated since:
\begin{equation}
\frac{\dd M}{\dd t} \sim - \sigma_{SB} T^2 \sim -M \, ,
\end{equation}
where the evaporation rate is given by Stefan-Boltzmann's law in 2d, and we used the fact that $T \sim \sqrt{M}$.} 
The exact solution for $\tau(t)$ is of the general form\footnote{Here $\alpha = \frac{24\pi}{c}\sqrt{\frac{E_0}{\kappa}}$. One can check that indeed $\tau(0)=0$, $\tau'(0)=1$, $\tau''(0)=0$ and $\tau'''(0)=-\frac{2\mu_0}{a\tau'(0)} = -2\kappa E_0$ as required to glue this solution to $\tau(t)=t$ for $t<0$ through an infalling pulse with magnitude $E_0$ using the boundary conditions for infalling pulses. 
These conditions fully fix all integration constants. } 
\bea
\label{exsolu}
\tau(t) \! \is \!  \frac{1}{\alpha^2} \int_0^t\!\! {\dd x\, \frac{1}{\bigl(I_1\left(\alpha \right)K_0\bigl(\alpha e^{-\frac{Ax}{2}}\bigr) + K_1\left(\alpha \right) I_0\bigl(\alpha e^{-\frac{Ax}{2}}\bigr)\spc \bigr)^2}} \, .
\eea
This solution has the property that $\tau'(+\infty) = 0$, implying that $\tau(+\infty) < +\infty$, thus Poincar\'e time does not flow forever in this solution.\footnote{The integrand behaves for large $x$ as $\sim 1/x^2$ which is integrable as $x\to+\infty$.}
The (quasi-static) Hawking temperature of the black hole as it evaporates equals
\begin{equation}
\label{qHawking}
T(t) = \frac{1}{\pi}\sqrt{\kappa {E(t)}} = \frac{1}{\pi}\sqrt{\kappa{E}}\exp\left(-\frac{A}{2} t\right) \, ,
\end{equation}
decaying at half the rate. 

We plot the exact solutions in Figure \ref{XevapA1}. One sees that as the evaporation rate increases, the profile approaches more and more the Poincar\'e profile. In the limiting case, the black hole evaporates instantaneously and the Poincar\'e time coordinate remains intact. For extremely low evaporation rates, one approaches the static black hole profile. Intermediate rates lead to postponing the Poincar\'e time $\tau$ at which this time coordinate stops flowing ($\partial_t \tau=0$). 
\begin{figure}[h]
\centering
\begin{minipage}{0.3\textwidth}
\centering
\includegraphics[width=\textwidth]{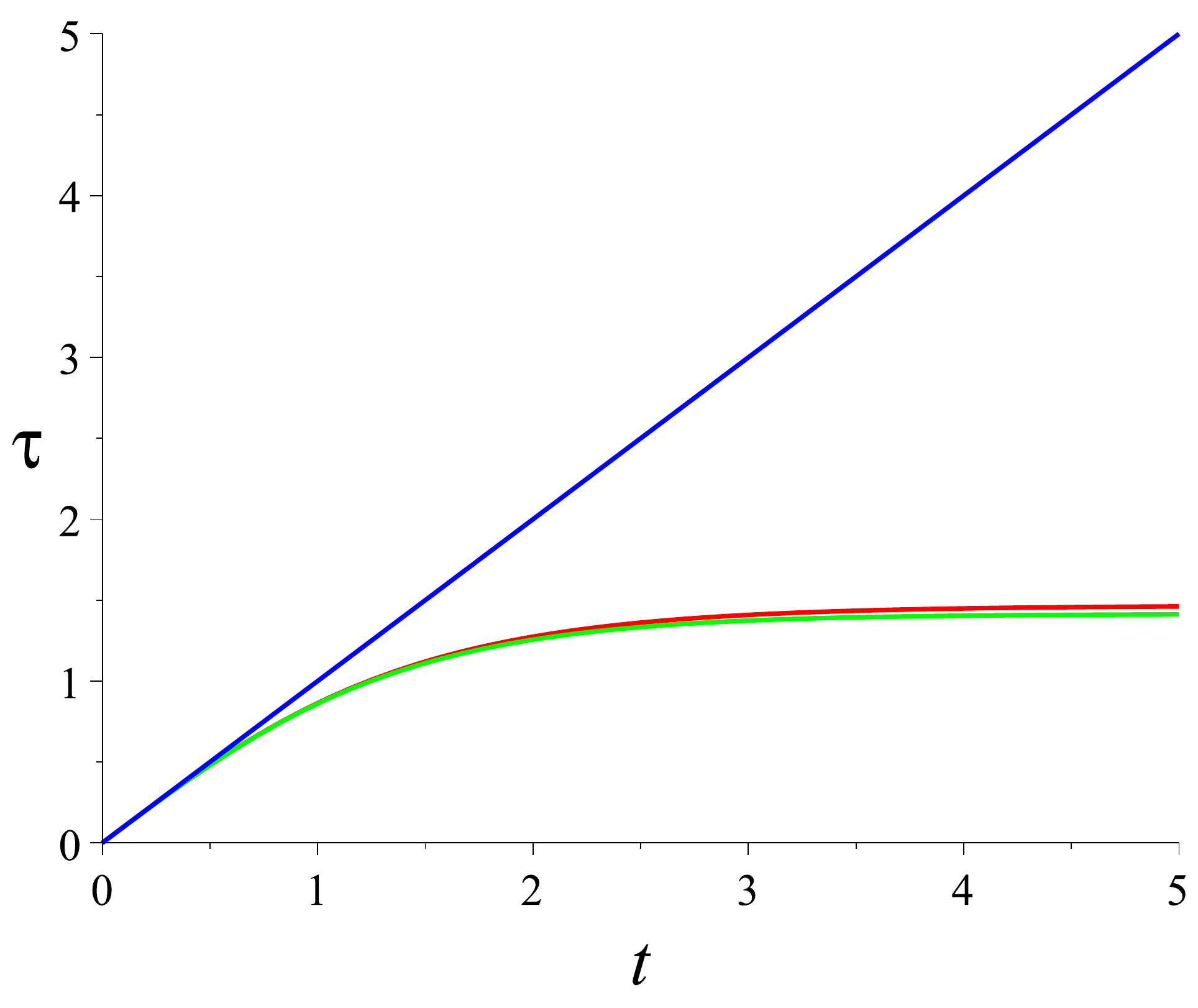}
\caption*{(a)}
\end{minipage}
\begin{minipage}{0.3\textwidth}
\centering
\includegraphics[width=\textwidth]{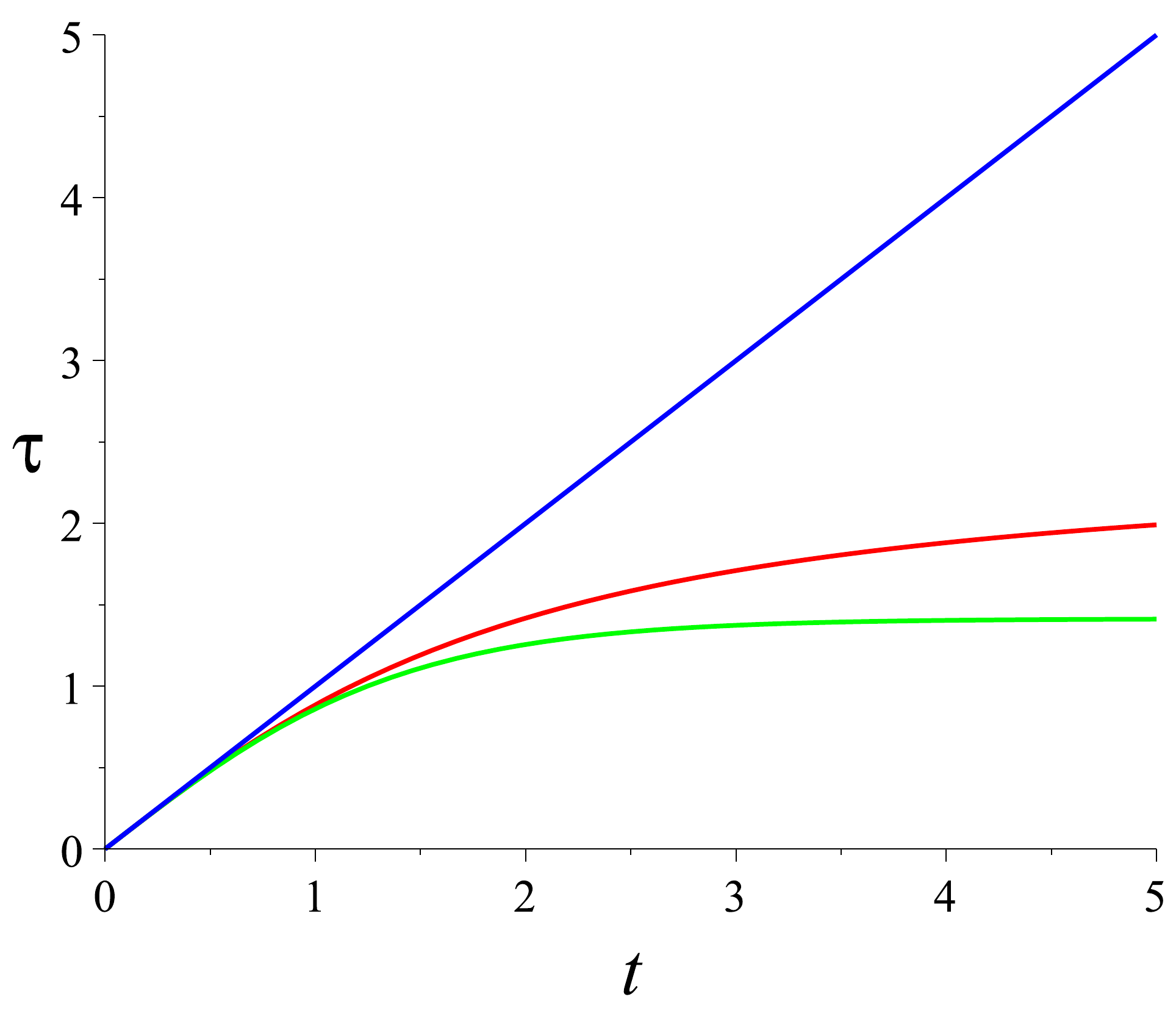}
\caption*{(b)}
\end{minipage}
\begin{minipage}{0.3\textwidth}
\centering
\includegraphics[width=\textwidth]{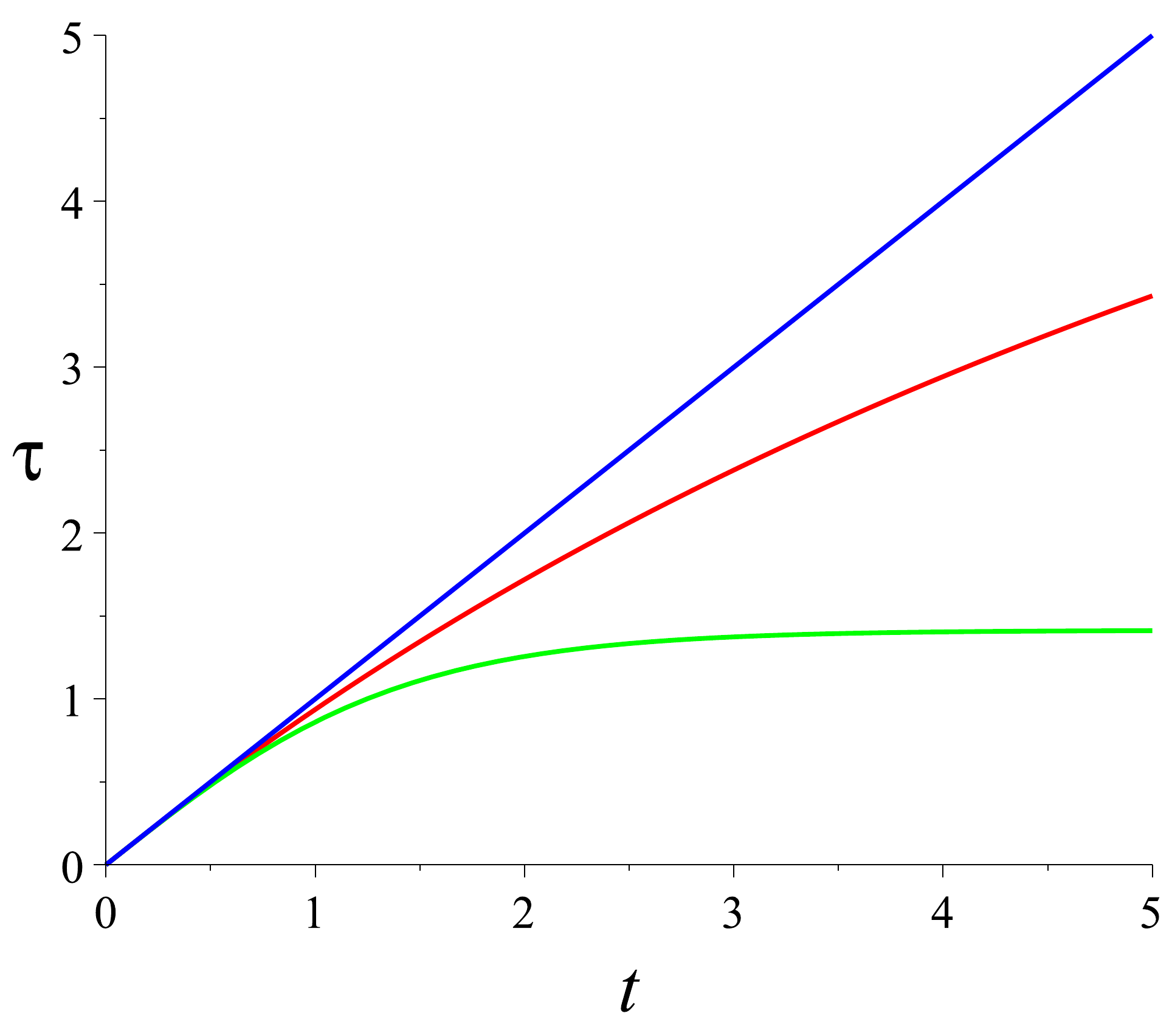}
\caption*{(b)}
\end{minipage}
\caption{$\tau$ as a function of $t$ for $2\kappa E_0=1$. The red curve represents the exact solution (\ref{exsolu}), the green curve represents the case where there is no evaporation, and the blue curve is the Poincar\'e curve for reference. (a) $A=0.1$. (b) $A=1$. (c) $A=5$.}
\label{XevapA1}
\end{figure}

\def\lll{L}
\section{Conclusion}
\label{concl}

In this work we set out to analyze more deeply the holographic features of the Almheiri-Polchinski dilaton gravity model.
The bulk dilaton gravity model contains a natural dynamical time variable $\tau(t)$ that dictates the relation between the time $t$ that a boundary observer experiences, relative to the Poincar\'e reference frame.  Within this flowing time frame, the holographic stress tensor expectation value has a particularly elegant form: as a Schwarzian derivative of the time reparametrization (\ref{bstress}). \\
We presented a Hamiltonian formulation of the boundary time variable, and uncovered an $SL(2,\mathbb{R})$ symmetry group, that acts via M\"obius transformations on $\tau(t)$. When coupling the dynamical boundary to the matter sector, an additional Virasoro structure of conserved charges emerges.

We then set out to analyze chaotic behavior in this model. Commutators of out-of-time local operators of the matter CFT experience maximal Lyapunov behavior, which can be generalized in the form of an exchange algebra, revealing the underlying shockwave interaction in the bulk.

Finally, we considered some quantum aspects of this model. Non-local effective actions can be constructed as generators of correlation functions. Also deformations of the model were studied, demonstrating that this model still has a lot more interesting features waiting to be uncovered. In the final section, we provided an example of a dynamically evaporating black hole, as described by the preferred time coordinate. An exponential decay was found, merely because the outgoing flux and the ADM energy are both given by Schwarzian derivatives.

There are several open ends to the story. A key question is whether the dilaton-gravity-matter system all by itself can be developed into a complete self-consistent quantum theory,
or whether it should be viewed as an effective bulk theory with a more fundamental description in terms of a holographic dual, given by some quantum many body system similar to the SYK model. 

Entropy considerations often give useful guidance. Entropy indeed appears to play an  interesting dynamical role in the AP model. From the formula (\ref{temperature}) of the Hawking temperature, we read off that the entropy and energy are related via
\bea
\label{entropy}
S \is 2\pi \sqrt{\frac{E}\kappa} \, = \, 2\pi \sqrt{\frac{a}{8\pi G} \, E } \, .
\eea
Is it possible to explain this formula via the counting of micro states?
Equation (\ref{entropy}) is of course reminiscent of a Cardy formula for a 2d CFT on a spatial circle of length $L$
\bea
S_{\rm cardy} \is 2\pi \sqrt{\frac{c}{6}\lll\spc E} \, .
\eea
The two formulas would match if we identify
\bea
\frac{a} {8\pi G} \, = \, \frac{c}{6}\,  \lll \, .
\eea
This equality hints that the integration constant $a$ should  perhaps be viewed as an effective IR cut-off for the bulk CFT. 

There appears to be an intimate relation between the dilaton and the entropy. As seen from (\ref{BHdilaton}), the value of the dilaton at the horizon equals 
$\Phi^2_{|\rm horizon} = 1 + a \sqrt{\kappa E}$. Comparing with (\ref{entropy}) gives as an analogue of the Bekenstein-Hawking formula 
\bea
S_{\rm bh}  \is \frac{1}{4G} \bigl(\Phi^2-1\bigr)_{|\rm horizon}  \, .
\eea
This identification has the following intriguing generalization, analogous to the Ryu-Takayanagi formula \cite{RT}. Consider equation (\ref{dilat}) for the dilaton profile in the bulk.
Suppose we set $T_{++}(s) = T_{--}(s)$, as would follow from a reflecting boundary condition at the AdS${}_2$ boundary. Equation (\ref{dilat}) then gives that the deviation $\delta \Phi^2$ of $\Phi^2$ due to the energy momentum flux is equal to
\bea
\label{dilats}
 \delta \Phi^{2}(x^+,x^-) \! \is\!  4G\, \delta S(x^+,x^-) \, ,
 \eea
 where
 \bea
 \delta S(x^+,x^-) \! \is \! 2\pi \int_{x^-}^{x^{+}}\!\!\!\!\!\dd s\, \frac{{(s - x^{+})(s - x^{-} )}}{x^+-x^-}\, T_{00}(s) \, .
\eea
This formula precisely matches with the first law of entanglement thermodynamics of a 2d CFT (see for example \cite{deBoer:2015kda}\cite{Asplund}), that expresses the change in the entanglement entropy of the interval $[x^+,x^-]$ due to an injection of energy momentum. 
This relation gives an encouraging  hint that it should be possible to reconstruct the complete bulk dynamics of the AP model from thermodynamic considerations. We leave this problem for future study.

\bigskip

\section*{Acknowledgements}
We thank Nele Callebaut, Juan Maldacena, Douglas Stanford, Grisha Tarnopolsky and Zhenbin Yang for valuable discussions and helpful comments. JE thanks Princeton University for the hospitality while this research was carried out. TM gratefully acknowledges financial support from Princeton University, the Fulbright program and a Fellowship of the Belgian American Educational Foundation. The research of HV is supported by NSF grant PHY-1314198.


\appendix

\section{Boundary dynamics in the global AdS${}_2$ frame}
\label{pulseinglobal}

In this section we briefly discuss the formulation of the boundary dynamics relative to the global frame.
The global coordinates $Y^\pm$ of AdS${}_2$ are obtained from the Poincar\'e coordinates $X^\pm$ by the transformation
\begin{equation}
X^+ = \tan Y^+, \quad X^- = \tan Y^- \, .
\end{equation}
The static global AdS${}_2$ form of the metric and dilaton read
\begin{align}
\dd s^{2} &= -\frac{4}{\sin^2(Y^+-Y^-)}\dd u\dd v\,, \\
\Phi^{2} &= 1 + a\coth (Y^+ -Y^-) \, ,
\end{align}
with $Y^+-Y^- \in \left[0,\pi\right]$ and $Y^+ + Y^- \in (-\infty,+\infty)$.

The boundary equation of motion, when starting in the global frame\footnote{The stress tensors are evaluated in the global frame here. We set $\tau = \tan y$ and start with $E=-\frac{1}{\kappa}$.} is given by
\begin{align}
\partial_{t}y &= 1 - \kappa \left(I_+(t) + I_-(t)\right) \, ,\\
I_+(t) &= \int_{y}^{+\infty}\dd s\sin^{2}(s - y)T_{++}(s) \, ,\\
I_-(t) &= \int_{-\infty}^{y}\dd s\sin^{2}(s - y)T_{--}(s) \,.
\end{align}
As an example of the use of this equation, suppose we send in a pulse directly in the global frame of the form
\begin{equation}
T_{--}= \omega \delta(Y^-)\,,
\end{equation}
then one obtains for respectively weak ($\kappa\omega < 1$) and strong pulses ($\kappa\omega >1$):
\begin{align}
\tan y &= \sqrt{\frac{1}{1 - \kappa\omega}}\tan(\sqrt{1-\kappa \omega}\, t) , \quad \kappa\omega < 1 \, , \\
\tan y &= \sqrt{\frac{1}{\kappa\omega-1}}\tanh(\sqrt{\kappa \omega-1}\, t), \quad \kappa\omega > 1\, .
\end{align}
Clearly, in the former case, no periodicity in imaginary time is generated, and this does \emph{not} represent a black hole; the infalling pulse is too weak. In the latter case, the pulse generates a black hole, by identifying $\kappa\omega - 1 = \kappa E $ where $E$ is the black hole mass.
This illustrates the fact that the global frame can be seen as being lower in energy by $-\frac{1}{\kappa}$ than the Poincar\'e patch, and this is easily confirmed by computing $\bigl\langle\, \hat{T}_{tt}\spc \bigr\rangle$.

The energy of the different spacetimes we discussed so far is illustrated in Figure \ref{energyscheme}.
As described in \cite{Balasubramanian:1999re} for the 3d case, this can be interpreted as the global frame having lower vacuum energy than the Poincar\'e patch. 
Furthermore, by starting in the global patch with $E = -\frac{a}{8\pi G}$ and letting the preferred frame evolve from there, one readily proves the consistency check:
\begin{equation}
\bigl\langle\, \hat{T}_{tt}\spc \bigr\rangle = -\frac{1}{2\kappa}\left\{\tan y,t\right\} = -\frac{1}{2\kappa}\{\tau,t\}\,.
\end{equation}
\begin{figure}[H]
\begin{center}
\medskip
\includegraphics[width=0.46\textwidth]{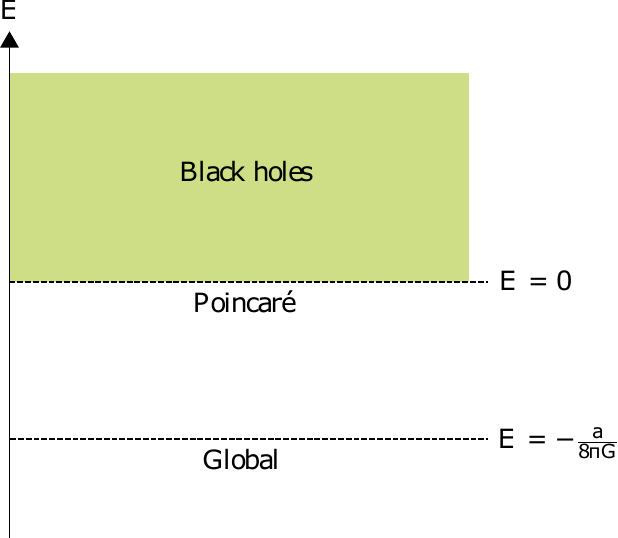}
\end{center}
\vspace{-2mm}
\caption{Energy of the spacetime in the family of coordinate patches related to Poincar\'e coordinates by having the same asymptotic form. The global frame is lower in energy than the Poincar\'e frame.}
\label{energyscheme}
\end{figure}

\section{Multi-pulse dynamics}
\label{multipulse2}
Next to the 3 standard frames we studied up to now (Poincar\'e, black hole and global), there are a host of new frames that can be obtained by sending in multiple pulses with time delays in between. These have the same (classical) boundary energy as the previous ones however. Since the boundary energy is a Schwarzian derivative (\ref{bstress}), these frames must be related by a M\"obius transform to the 3 main frames.

It is even possible to set up a transfer matrix framework to write the most general multi-pulse solution, relating $\tau(t)$ where all pulses are sent at the same time, to that in which arbitrary time delays are present by an $SL(2,\mathbb{R})$ transformation.

Denote
\begin{equation}
\mathbf{A}(\Delta; E)=
\begin{pmatrix}
\cosh\left(\sqrt{\kappa E}\Delta\right) & \frac{1}{\sqrt{\kappa E}}\sinh\left(\sqrt{\kappa E}\Delta\right) \\
\sqrt{\kappa E}\sinh\left(\sqrt{\kappa E}\Delta\right)& \cosh\left(\sqrt{\kappa E}\Delta\right) 
\end{pmatrix} \, ,
\end{equation}
which is an $SL(2,\mathbb{R})$ matrix. Then one can write for $n$ pulses ($E_1 \hdots E_n$) separated by $\Delta_1 \hdots \Delta_{n-1}$, the M\"obius transformation linking the case where all energy falls in at once
\begin{equation}
\tau(t) = \frac{1}{\sqrt{\kappa \sum_j E_j}}\tanh\left(\sqrt{\kappa \sum_j E_j}t\right) \, ,
\end{equation}
to the solution $\tilde{\tau}(t)$ which includes the time delays as a concatenation of these M\"obius transforms:
\begin{align}
\tilde{\tau}(t) &= \mathbf{A}(\Delta_1;E_1) \mathbf{A}(\Delta_2-\Delta_1;E_1+E_2) \hdots \mathbf{A}\bigl(\Delta_i-\Delta_{i-1};\sum_{j=1}^{i} E_j\bigr) \nonumber \\
&\quad \quad \times \hdots \mathbf{A}\bigl(\Delta_{n-1}-\Delta_{n-2};\sum_{j=1}^{n-1} E_j\bigr)\mathbf{A}\bigl(-\Delta_{n-1};\sum_{j=1}^{n} E_j\bigr) \cdot \tau(t) \, .
\end{align}

Frames related by M\"obius transforms exhibit unitarily equivalent QFT constructions and hence for quantum phenomena, it is irrelevant how precisely one obtains the classical solution to start with, as one would also conclude intuitively.

\end{document}